\documentclass[twocolumn,preprintnumbers,amsmath,amssymb,nofootinbib
,superscriptaddress,showpacs]{revtex4}
\usepackage{hyperref}
\usepackage{graphicx}
\usepackage{color}

\newcommand{\be}{\begin{equation}}  
\newcommand{\ee}{\end{equation}}  
\newcommand{\bear}{\begin{eqnarray}}  
\newcommand{\eear}{\end{eqnarray}}  
\newcommand{\ba}{\begin{array}}  
\newcommand{\ea}{\end{array}}

\newskip\humongous \humongous=0pt plus 1000pt minus 1000pt

\newif\ifdtup

\def\oldreffmt#1{\rlap{[#1]} \hbox to 2\parindent{}}

\def\figfmt#1{\rlap{Figure {#1}} \hbox to 1in{}}  
  
\def\ie{\hbox{\it i.e.}{}}	  
\def\eg{\hbox{\it e.g.}{}}	  
\def\etal{\hbox{\it et al.}}  
  
\def\slash#1{#1\!\!\!/\!\,\,}  
\def\beq{\begin{equation}}  
\def\eeq{\end{equation}}  
\def\bea{\begin{eqnarray}}  
\def\eea{\end{eqnarray}}  
\def\half{\frac{1}{2}}  
  
\def\bq{\begin{quote}}  
\def\eq{\end{quote}}

\def\half{\frac{1}{2}}       

\def \etal {{\it et al.}\ }  
\relax  

\newdimen\tdim  
\tdim=\unitlength  
\def\bar{\overline}

\begin{document}

\preprint{FERMILAB-PUB-15-350-T}

{\title{Axion Induced Oscillating Electric Dipole Moment of the Electron 
}

\author{Christopher T. Hill}
\email{hill@fnal.gov}
\affiliation{Fermi National Accelerator Laboratory\\
P.O. Box 500, Batavia, Illinois 60510, USA}

\date{\today}

\begin{abstract}
A cosmic axion, via the electromagnetic anomaly, induces an oscillating electric dipole for the electron
of frequency $m_a$ and strength $\sim$(few)$\times 10^{-32}$ e-cm,
 two orders of magnitude above the nucleon, and 
within a few orders of magnitude of the present standard model constant limit.
We give a detailed study of this phenomenon via the interaction of the cosmic axion, 
through the electromagnetic anomaly, with particular
emphasis on the decoupling limit of the axion, $\partial_t a(t)\propto m_a \rightarrow 0$.
The analysis is subtle, and we find the general form of the action involves a local contact interaction
and a nonlocal contribution, analogous to the ``transverse current'' in QED, that enforces the decoupling limit.  
We carefully derive the effective action in the
Pauli-Schroedinger non-relativistic formalism, and in   Georgi's heavy quark formalism 
adapted to the ``heavy electron'' ($m_e>>m_a$). We compute the
electric dipole radiation emitted by free electrons, magnets and currents, immersed in the cosmic axion field,
and discuss experimental configurations that
may yield a detectable signal.

\end{abstract}


\maketitle

\section{Introduction}

In the present paper we give a detailed study of the interaction of the cosmic axion 
through the electromagnetic anomaly,  with  the 
magnetic dipole field of an electron. For an oscillating cosmic axion
field, we show that the electron acquires
 an  effective oscillating electric dipole moment (OEDM).  Our detailed analysis is subtle,
and amplifies the result of our previous note \cite{Hill1}. 
The analysis is perturbative, and is treated both classically and quantum mechanically. 
Not surprisingly, we find that the OEDM displays subtleties that are shared with the axial anomaly itself. 

We'll assume that the axion fills the vacuum as a classical coherent field, 
oscillating in a given frame with frequency $m_a$, and may be
associated with dark matter, as per the model in \cite{Ipser}. We will be interested
in the effect upon magnetic objects that are essentially at rest relative to the axion cosmic rest-frame.
The axion field is designated as $\theta(t) = a(t)/f_a$ where $a$ is the canonically normalized
axion field and $f_a$ the decay constant. The cosmic axion field is then $\theta(t) = \theta_0 \cos(m_a t)$ where
$m_a\sim m_\pi f_\pi/f_a$ is the axion mass. 

A simple hand-waving argument can be given for the existence of induced OEDM's 
arising from magnetic moments immersed in a cosmic axion field.
Witten showed  that a $\theta$-angle
in  QED will cause magnetic monopoles to acquire electric charges proportional
to $\theta$, \ie, they become ``dyons'' \cite{Witten, Sikivie0}.  If we thus consider a 
pair of dyon and anti-dyon, separated
in space by a very small distance, we will have both a magnetic and an electric dipole moment
where the electric dipole moment is
equal to the magnetic moment times $\theta$.  
At very large distances we cannot, for all practical purposes, discern the
presence or otherwise of the underlying magnetic monopoles, but the electric and magnetic dipole fields persist.
A cosmic axion filling space is essentially an oscillating $\theta$-angle, and we might expect by
this argument, therefore, an OEDM proportional to $\theta(t)$. 

An OEDM
is indeed induced for the electron by the Feynman diagram of Fig.(1) where the solid dot vertex is
the  anomalous coupling of the axion to electromagnetic fields.
{
Our result can be written as an effective interaction 
for the non-relativistic electron in the zero electron--recoil
limit, with a {\em time dependent electric
field} (\eg, radiation or a cavity mode) as:
\bea
\label{edm00}
&& \int d^4x \;2g_{a}\theta(t)\; \mu_{Bohr}\;
\chi^\dagger \frac{\overrightarrow{\sigma}}{2} \chi(x) \cdot \overrightarrow{E}(x,t)
\eea
where $g_{a}$ is the axion-$\gamma$-$\gamma$ anomaly coefficient  (defined below),
$ \mu_{Bohr}$ the Bohr magneton, and $\overrightarrow{E}(x,t)$
is an external radiative electric field.  
The induced effective oscillating electric dipole moment is
proportional to  the magnetic moment, $d_e\approx 2g_{a}\theta_0\cos(m_at)\mu_{Bohr}
\approx 3.2\times 10^{-32}(g_{a}/10^{-3})\cos(m_at)$
e-cm. We use the term, ``effective,'' because this arises at the level of a one-particle reducible
Feynman diagram.  Note that Eq.(1)
is the limit of the action in the case of source-free, time dependent
(radiation) fields, and one cannot naively take the limit $\theta(t)
\rightarrow$ 
(constant) without including the nonlocal terms in
the full action, as in Eqs.(29), (30).

Previously, an OEDM has only been considered to be specific to
baryons. It arises, not by the electromagnetic 
anomaly,  but rather directly
via the QCD-induced axion potential.  
The magnetic moment of the electron is much larger
than that of the nucleon, and hence 
the axion-induced oscillating electric dipole moment is almost three orders
of magnitude larger than that of the nucleon.
Since the current best limit upon any DC elementary particle EDM is that  of the electron, 
 of order $d_e\leq 8.7\times 10^{-29}$ e-cm, \cite{ACME}, 
the electron may provide a promising place
to search for an oscillating EDM. 

While the OEDM appears above to be
proportional to $\theta(t)$, there is, however, a catch:   
In the limit that $\partial_t \theta(t)\rightarrow 0 $ (decoupling limit)
the perturbative Feynman diagrams involving the anomaly must  vanish.
But how do we reconcile decoupling, \ie, derivative coupling of the axion, from 
a hard dependence upon $\theta(t)$? 

The decoupling of the axion
at zero mass is subtle. 
We will see presently that a nontrivial nonlocal term is generated by Fig.(1) that enforces the decoupling. 
This nonlocality is remniscent of the ``transverse current'' that arises in
radiation gauge quantization of QED (see, \eg, section 6.3 of ref.\cite{Jackson}).  
The nonlocal term insures that the action,
$S(\theta, F_{\mu\nu}, ...)$ can be brought to the form
$S'(\partial_\mu\theta, A_{\nu}, ...)+$(total divergence), where $ F_{\mu\nu}=\partial_{[\mu} A_{\nu ]}$.
This is a subtle property shared with the anomaly itself whereby
the manifestly gauge invariant form of $S$ does not display the axion derivative coupling.
However, upon integration by parts we can display the derivative coupling
of the axion, as in $S'$, while relinquishing manifest gauge invariance.
Since these actions are equivalent up to a total divergence, both
the shift symmetry of the axion and the gauge invariance of QED remain valid in
perturbation theory.  Displaying the action as in $S$ and taking the zero
recoil limit of the electron, which is kinematically valid in the $m_e>>m_{a}$ limit,
we obtain eq.(\ref{edm00}). The price we pay for this symmetry is the
nonlocality of the effective EDM action of the electron. The structure of the
action is, however, determined completely by this symmetry, as we discuss in Section(III).

Indeed,   nonlocality 
arises even in the familiar case of
a classical axion induced RF cavity mode. There, the induced electric field in the cavity,
$\overrightarrow{E}(t)$ satisfies a similar condition,
$\overrightarrow{E}(t)=\overrightarrow{E}(t_0)+c(\theta(t) - \theta(t_0))$.
This happens simply because $\overrightarrow{E}(t)$ is governed by a first order
inhomogeneous differential Maxwell equation and requires  a boundary condition.
As we'll see, the particular solution of Maxwell's equations for an induced electric field
in a static background magnetic field is of fundamental importance
in axion electrodynamics and drives most of the interesting phenomena.
Modulo this subtlety, the explicit calculation of the Feynman diagram as in Fig.(1)
nontrivially confirms the argument based upon Witten's dyons. The full calculations
 simultaneously provides consistency with decoupling via the nonlocal term.
}

\vskip 0.2in

We begin by giving a detailed derivation of the effective action
of the electron OEDM in Section(II). We consider both the non-relativistic Pauli-Schroedinger formalism
for a resting electron,
and also  Georgi's covariant heavy quark formalism for electron of 4-velocity $v_\mu$ \cite{georgi}. 
The latter formalism is adapted to the 
electron, which may be viewed as ultra-heavy in comparison to $m_a$, and shows that
the resulting interaction is of the form $\propto \theta(t)\bar{\psi}_v\sigma_{\mu\nu}\gamma^5\psi_v F^{\mu\nu}$
for  $\psi_v=(1+\slash{v})\psi/2$, with 4-velocity $v_\mu$.
The results are consistent, and reveal the full effective action with the nonlocal term.  

In Section III.A we
show that the structure of the action with the nonlocal term is completely determined by the axion decoupling, \ie,
by the  shift symmetry, $a/f_a\rightarrow a/f_a + \phi$, which is maintained in perturbation theory.
While the physical effective action of the OEDM is consistent with 
the  $a/f_a\rightarrow a/f_a + \phi$ symmetry, we emphasize that there are no additional suppressions 
involving higher powers of $m_a$, \ie, our OEDM physics is on par with the induced oscillating 
electric field in an RF cavity experiment. 
In Section III.(B,C), we observe how  this nonlocality  arises 
 in well-known solutions to, \eg, the RF cavity experiments. 

 To further probe this phenomenon, we show explicitly in Section IV that the classical Maxwell equations for a localized magnetic dipole, such as an electron in free space, leads to the emission of electric dipole radiation, \ie,
the classical radiation field from a stationary electron  is that of a Hertzian electric dipole radiator.
{
The classical calculation is compared to the quantum calculation, and they are found to be consistent.
Large magnetic fields imbedded in conductors likewise provide a source for such 
axion induced electric dipole radiation.

In Section V we consider a possible experimental configuration for detection of
this radiation based upon an array.  This is a broadband
simple radiator, and can be viewed as an array of high field magnets,
or as a planar slab of conductor with
a large magnetic field imbedded in the plane of the conductor. 
This can produce power output of upwards of order $\sim 10^{-24}$ watts
and appears to be detectable radiometrically.
The main advantage over RF cavity
experiments is that  broad-band radiators  do not
require resonant tuning. } We are encouraged by the simple estimates
of the signal integration that we provide that this
may lead to detactability, even in the challenging range
$10^{12} \geq f_a \geq 10^{10}$ GeV, or short axion wavelength.
We note that there are several  papers that touch on these and related ideas, \eg, 
\cite{Krauss}, \cite{Raffelt}, and \cite{Ringwald}.

\vskip 0.2in

Let us  recall some basic concepts. 
The axion is a hypothetical, low--mass pseudo-Nambu-Goldstone boson (PNGB) that offers 
a  solution to the strong CP problem of the standard model, and
simultaneously  provides a compelling dark matter candidate.
The expected mass scale of the axion is $m_{a} \approx {m_{\pi }^{2}}/{f_a}$ where
typical expected values of the decay constant $f_a$ range from $\sim 10^{10} $ GeV upwards
\cite{PQ1,Weinberg1,Wilczek1}.    

The axion is  expected to have an anomalous coupling to the electromagnetic field $
\overrightarrow{E}\cdot $ $\overrightarrow{B},$ taking the form:
\beq
\label{zero}
\frac{g_a}{4}\int d^4 x\;\left( \frac{a}{f_a}\right) F_{\mu\nu}\widetilde{F}^{\mu\nu}=
-g_a\int d^4 x\;\left( \frac{a}{f_a}\right) \overrightarrow{E}
\cdot \overrightarrow{B}
\eeq
where $\widetilde{F}_{\mu\nu}=(1/2)\epsilon_{\mu\nu\rho\sigma}{F}^{\rho\sigma}$, 
and  $g_a $  is the dimensionless
anomaly coefficient. In various models we have \cite{PDG,DFSZ,KSVZ}}: 
\bea
g_a & \approx & 8.3\times 10^{-4}\qquad \;\;\; \makebox{DFSZ }
\nonumber \\
g_a &\approx & -2.3\times 10^{-3}\qquad \makebox{KSVZ }
\eea
In making quantitative estimates
in Section V. we will use the KSVZ result
(see Section V, in particular Table I,
 for a listing of our preferred parameters).

Most strategies for detecting the cosmic axion
exploit the electromagnetic anomaly \cite{Bj,Sikivie1,Graham}
together with the assumption of a coherent galactic dark-matter background field \cite{Ipser},  
$ {a}/{f_a}\equiv \theta(t) =$\ $\theta_{0}\cos (m_at)$.  
In typical RF cavity experiments such as ADMX, one applies a large external 
constant magnetic field to the cavity,
$\overrightarrow{B}_0$  (it suffices to apply this field only
in the conducting walls of the cavity),
and the anomalous coupling to $\theta(t)$ induces an oscillating electromagnetic response field,
$\overrightarrow{E}_r$ and $\overrightarrow{B}_r$.
The ``cavity modes'' can become excited,
which can  generate a resonant signal in the cavity. This
offers the possibility of both detecting the existence of the axion
and simultaneously establishing that it is a significant component of dark-matter.
We briefly discuss, for sake of comparison, 
the energetics of RF in Section III.C, and quantitatively in Section V.

Recently several authors have considered alternative modes
of axion detection \cite{budkher, stadnik, graham2}, in particular, 
the possibility of observing an OEDM for the nucleons. Indeed,  
a small oscillating electric dipole moment for the nucleons is predicted with a frequency $m_a$
given by $d_N\sim 10^{-16}\theta(t)\approx 3.67\times 10^{-35}\cos(m_at)$ e-cm \cite{budkher}.  
Thus far, this effect has only been considered to be specific to
baryons. It arises, not by the electromagnetic anomaly $\propto g_a$,  but rather directly
via the QCD-induced axion potential.

\begin{figure}[t]
\vspace{4.5cm}
\includegraphics{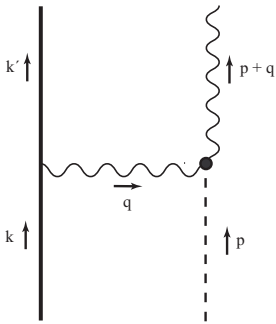}
\vspace{-0.0cm}
\caption[]{ Feynman diagram for axion induced electric dipole moment.  Solid
dot is the axion anomaly interaction, $\theta\overrightarrow{E}\cdot\overrightarrow{B}$. The dashed line
is the incoming axion field, $\theta$. 
The solid line is the  electron, where the electron-photon vertex
is the magnetic dipole moment operator of the electron. }
\label{figone}
\end{figure}

The axion induced magnetic moment of the electron is about two orders
of magnitude larger than that of the nucleon.
Since the current best limit upon any DC elementary particle EDM is that  of the electron, 
 of order $d_e\leq 8.7\times 10^{-29}$ e-cm, \cite{ACME}, 
the electron may be a promising place
to search for an oscillating EDM and the axion itself.

\newpage

\section{ Feynman Diagram Analysis of Induced OEDM of the Electron}

\subsection{ Nonrelativistic Pauli-Schroedinger Action}

We compute the axion induced OEDM 
of the electron. We go to the electron rest frame which we assume to also be the
rest frame of the cosmic axion field, \ie, the frame
in which the axion field oscillates in time with
no spatial dependence, $a/f =\theta_0\cos(m_at)$. We use the Pauli-Schroedinger
formalism, then follow with an analysis in Georgi's heavy fermion formalism. 

The Pauli-Schroedinger Lagrangian
 is, of course, the first order Dirac action in an expansion in $1/m_e$, and
takes the form:
\bea
\label{PS1}
& & \frac{1}{2m}\psi ^{\dagger }\overrightarrow{
\sigma }\cdot \left( i\overrightarrow{\partial }-e\overrightarrow{A}\right) 
\overrightarrow{\sigma }\cdot \left( i\overrightarrow{\partial }-e
\overrightarrow{A}\right) \psi 
\eea
where $\psi$ is a two-component spinor.
Eq.(\ref{PS1}) contains the term:
\bea
\label{magmom}
& &  -\frac{1}{2m}\psi ^{\dagger }\overrightarrow{\sigma }\cdot \left( i
\overrightarrow{\partial }\right) \overrightarrow{\sigma }\cdot \left( e
\overrightarrow{A}\right) \psi
\supset -\frac{ie}{2m}\psi ^{\dagger }\epsilon _{ijk}\sigma _{k}\psi 
\partial _{i}A_{j}
\nonumber \\
\eea
having used  $\sigma _{i}\sigma _{j}=\delta
_{ij}+i\epsilon _{ijk}\sigma _{k}$. This is
the standard magnetic dipole interaction, $ -\frac{ige}{2m} \psi^{\dagger }\frac{\vec{\sigma}}{2}\psi \cdot \vec{B}
$  with $g=2$.

Consider the time-ordered product of $i\times$ the
magnetic dipole action with $i\times$ the  axion anomaly action: 
\bea
& & \!\!\!\!\!\!\!\!\!\!\!\!\!\!\!\!\!\!
-(i)^2T\;\int d^{4}x\frac{ie}{2m}\psi ^{\dagger }\epsilon _{ijk}\partial
_{i}A_{j}\sigma _{k}\psi\;\;\; 
\nonumber \\
& &  \qquad \qquad \times {\tiny \frac{g_a}{2}\tiny }\int d^{4}y\theta \epsilon
^{\alpha \beta \gamma \rho }\partial _{\alpha }A_{\beta }\partial _{\gamma
}A_{\rho }
\eea
Assuming  $\theta$ has only time dependence, and
integrating the anomaly by parts in time we have
\bea
\label{ssix}
iT \frac{eg_a}{4m} \!\! \int \! d^{4}x\: \psi ^{\dagger }\epsilon _{ijk}\sigma _{k}\psi 
\partial _{i}A_{j} \!\! \int \! d^{4}y  (\partial _{0}\theta)
\epsilon_{lmn}A_{l}\partial _{m}A_{n}
\nonumber \\
& & 
\eea
Note that
this  forces us into radiation gauge,
as a Coulomb term would be $\sim \theta \epsilon_{lmn}\partial _{l}\phi \partial_{m}A_{n}$ 
and upon integration by parts we would have $\sim -\partial _{l}\theta \epsilon_{lmn}\phi \partial _{m}A_{n}$,
which vanishes by the hypothesis that the axion field depends only upon time.

Contracting the vector potential from the magnetic dipole interaction with
either vector potential in the anomaly yields:
\bea
\label{PS2}
&&\!\!\!\!\!\!\!\!
\frac{eg_a}{2m} 
\int d^{4}y\int d^{4}x\;\psi(x) ^{\dagger }\epsilon
_{ijk}\sigma _{k}\psi(x)
\nonumber 
\\
&&\qquad \qquad \times \partial _{i} G(x-y)\partial _{0}\theta(y^0) 
\epsilon_{jmn}\partial _{m}A_{n}(y)
\eea
where we've included a $2\times $ combinatorial factor.
Here we use the covariant Feynman propagator of the photon, 
\bea
\label{Green}
i\left(g_{\mu\nu}-\lambda \frac{\partial_\mu\partial_\nu}{\partial^2}\right) G(x-y)
& \rightarrow &-i\delta_{ij}G(x-y);
\nonumber \\
G(x-y)  =  -\int \frac{d^4k}{(2\pi)^4} && \!\!\!\!\!\!\!\! \frac{e^{ik\cdot (x-y)}}{k^2 + i\epsilon }
\eea 
since only the spatial terms
are relevant and the gauge dependent terms, 
$\propto \lambda $,
are seen to cancel owing to antisymmetries (see ref.\cite{Hill1}). 
Our sign conventions and normalization are consistent with ref.\cite{BjDrell}, where we have
$\partial^2 G(x-y)=\delta^4(x-y)$.
 
Using, $\epsilon _{ijk}\epsilon _{jmn}=-\delta
_{im}\delta _{kn}+\delta _{in}\delta _{km}$, we can write:
\bea
&& =-\frac{eg_a}{2m}\int d^{4}y\int d^{4}x \; \psi ^{\dagger }\sigma
_{k}\psi (x)
\nonumber \\
&& \qquad \qquad \times \; \partial _{i} G(x-y) \partial _{0}\theta(y^0) \left( \partial
_{i}A_{k}(y)-\partial _{k}A_{i}(y)\right)
\nonumber \\
\eea
Note that the result of eq.(\ref{PS2}) is manifestly proportional to $\partial_{0}\theta$
and therefore the axion is derivatively coupled up a total divergence.  However, when
displayed in this form we do not see the manifest gauge invariance.  This issue
is addressed in detail in Section III.

We assume that the electron is in a common rest frame with
the axion. 
To a good approximation the electron operator is static, \ie,
$\partial _{0}(\psi ^{\dagger }\sigma _{k}\psi ) =0+O(1/m)$.
since the electron mass is enormous compared to the relevant
momentum exchange, $m_e >> m_a$.  The electron can therefore
absorb this tiny  momentum without appreciably changing its energy.

We can therefore integrate the time derivative by parts, 
and drop  $\partial_{y^0}G(x-y)$ terms.
We can also integrate the spatial derivative $ \partial_{y_k}$  by parts, and
note the translational invariance of  the propagator implies,
$ \partial_{y}  G(x-y) = -\partial_{x} G(x-y)$.
This yields:
\bea
&& =\frac{eg_a}{2m}\int d^{4}y\int d^{4}x\; \psi ^{\dagger }\sigma _{k}\psi (x)
\nabla_x ^{2}G(x-y)\theta \left( \partial _{0}A_{k}(y)\right)
\nonumber \\
&&
 -\frac{eg_a}{2m}%
\int d^{4}y\int d^{4}x\; \psi ^{\dagger }\sigma _{k}\psi(x)
\partial _{k}G(x-y)\theta \left( \partial
_{i}\partial _{0}A_{i}(y)\right) 
\nonumber
\\
\eea
where $\nabla^{2}=\partial_k\partial_k$.
Using the definition of the electric field in radiation gauge, 
 $E_i = -\partial_0 A_i$, the static
electron field permits us to integrate over $x^0$,  $\int dx^0\; \nabla ^{2}G(x-y)=\delta^3(x-y)$.
We obtain the effective action (we have removed the $i$):
\bea
\label{finalPS}
&& =\frac{ieg_a}{2m}\left[\int d^{4}x\; \psi ^{\dagger }\sigma _{k}\psi \; \theta (t)E_{k}(x)\right.
\nonumber 
\\
&& \;\;\; \left. +\int d^{4}y\int d^{4}x\;\partial _{k}\psi ^{\dagger }\sigma
_{k}\psi(x) \ G(x-y)\theta (y^{0})\partial _{i}E_{i}(y)\right]
\nonumber \\
\eea
The result we have obtained contains the local contact interaction,
which is a conventional electric dipole form proportional to $\theta(t)$.
It also contains a nonlocal component. 
If we choose to immerse the electron in  a source free
field, such as an RF cavity mode or light, then $\partial _{i}E_{i}(y)=0$
and only the contact term remains.

We have written this expression in
a manifestly gauge invariant form, and therefore we display the
axion field without a derivative. Integration by parts, as in 
the general argument of Section III, shows that the axion is
derivatively coupled. Our interaction behaves like the anomaly
itself in its display of manifest gauge invariance
 vs manifest shift invariance..

\subsection{Georgi's Heavy Quark Effective Theory Applied
to the Electron}

The previous analysis relied upon the non-recoil approximation of
the electron, which is justified since $m_e>>m_a$. 
This limit implies that the $4$-velocity of the electron
is approximately conserved in its interactions with the comoving axion
field. 
A more ``covariant'' representation can be derived, however, by
using Georgi's heavy quark effective field theory formalism \cite{georgi}.
 We  obtain the
same basic structure as in the Pauli-Schroedinger case, but
formulae involving the electric dipole moment will
become the more familiar relativistic forms.

We begin by defining  heavy electron Dirac fields with fixed
4-velocity, $v_\mu$:
\beq
\psi \rightarrow \left( \frac{1+\slash{v}}{2}\right) \psi =\psi
_{v}\qquad \overline{\psi }\rightarrow \overline{\psi }\left( \frac{1+%
\slash{v}}{2}\right) =\overline{\psi }_{v}
\eeq
The Dirac magnetic moment operator then takes the form:
\beq
-\frac{ie}{4m}\overline{\psi }_{v}\sigma _{\mu \nu }\psi _{v}F^{\mu \nu
}=-\frac{ie}{2m}\overline{\psi }_{v}\left( \gamma _{5}\epsilon _{\alpha \beta \rho \gamma }\gamma ^{\rho
}v^{\gamma }\right) \psi
_{v}\partial ^{\alpha }A^{\beta }
\eeq
where  the axion field is $\theta=\theta_0\cos(m_a v_\mu x^\mu)$.

We again compute the time ordered product
of $i\times$ the
magnetic dipole action with $i\times$ the  axion anomaly action: 
\bea
\label{fourteen}
&& -(i)^2T\; \frac{ieg_a}{4m}\int d^{4}x\overline{\psi }_{v}(x)\left( \gamma
_{5}\epsilon _{\alpha \beta \rho \gamma }\gamma ^{\rho }v^{\gamma }\right)
\psi _{v}(x)\partial ^{\alpha }A^{\beta }(x)\; 
\nonumber \\ 
&& \qquad \times  \int d^{4}y
\theta (y)\epsilon ^{\eta \kappa \omega \rho }\partial _{\eta }A_{\kappa
}(y)\partial _{\omega }A_{\rho }(y)
\nonumber \\
&& =-\frac{eg_a}{2m}\int d^{4}y\int d^{4}x\overline{\psi }_{v}(x)\left(
\gamma _{5}\epsilon _{\alpha \beta \rho \gamma }\gamma ^{\rho }v^{\gamma
}\right) \psi _{v}(x)
\nonumber \\
&&\qquad \;\;\; \times \partial^{\alpha }G(x-y) \left( \partial
_{\eta }\theta(y)  \epsilon ^{\eta \beta \omega \rho }\partial _{\omega
}A_{\rho }(y)\right) 
\eea
Note again that propagator gauge terms do not contribute owing the the $\epsilon$-symbols. 
We now relabel, permute, and use the identities:
\beq 
\epsilon _{\alpha \beta \gamma \rho }\epsilon ^{\alpha \eta
\omega \delta }=-\left( g_{\beta }^{\eta }g_{\gamma }^{\omega }g_{\rho
}^{\delta }+(-1)^p(\makebox{permutations}) \right) 
\eeq
and:
\bea
-i\overline{\psi }_{v}\gamma
^{5}\sigma _{\alpha \beta }{\psi }_{v}
=\overline{\psi }_{v} \left( \gamma
^{5}\gamma _{\alpha }v_{\beta }-\gamma ^{5}\gamma _{\beta }v_{\alpha
}\right){\psi }_{v} 
\eea
Note that we can
integrate by parts and rearrange, to obtain:
\bea
\label{georgirep1}
& &  \!\!\!\!\!\!    \!\!\!\!\!\!
=\frac{ieg_a}{2m}\int d^{4}y\int d^{4}x\;
\overline{\psi }_{v}\gamma_{5}\sigma^{\rho \gamma }\psi_{v}
\left[
\partial^{2}G(x-y)(\partial_{\rho }\theta) A_{\gamma}
\right.
\nonumber \\
& &\left. \qquad
+ \partial^{\beta}\partial_{\gamma }G(x-y) \left( (\partial_{\beta }\theta)A_{\rho } - (\partial_{\rho }\theta) A_{\beta }
\right)\right] 
\eea
This form displays the manifest derivative coupling of the axion,
but not manifest gauge invariance.

Alternatively, we can  manipulate eq.(\ref{fourteen}),
using $\partial^2 G(x) =\delta^4(x)$, and multiplying
by $-i$ to obtain the action: 
\bea
\label{georgirep}
& & \!\!\!\!\!\!\!\!\!\! \!\!\!\!\!\!\!\!\!\!  = -\frac{eg_a}{4m}\int d^{4}x\; \overline{\psi }_{v}\left(
\gamma _{5}\sigma _{\rho \gamma }\right) \psi _{v} \theta F^{\rho \gamma} 
\nonumber \\
& & 
\!\!\!\!\!\!\!\!\!\! \!\!\!\!\!
-\;\frac{eg_a}{2m}\int\! d^{4}y\int \! d^{4}x\;\partial^\rho (\overline{\psi }_{v} 
\left( \gamma _{5}\sigma_{ \rho\gamma }\right) \psi _{v}(x))
\nonumber \\
&& \qquad\qquad\;\;\; \times
 G(x-y)\partial_{\beta}[\theta(y)F^{\gamma\beta}(y)] 
\eea
Here we have displayed manifest gauge invariance, while the derivative
coupling of the axion is not manifest.  We emphasize that eq.(\ref{georgirep})
and  eq.(\ref{georgirep1}) are equivalent up to surface terms.
Note that in eq.(\ref{georgirep}) we see  the familiar covariant EDM operators in
the heavy quark fields: $\overline{\psi }_{v}\left(
\gamma _{5}\sigma _{\rho \gamma }\right) \psi _{v}(x) $ .

To compare the Georgi calculation result to the Pauli-Schroedinger case we 
go to rest frame $v_{i}\rightarrow 0$, and carefully note
sign conventions in ref.\cite{BjDrell}:
\bea
&& -\frac{eg_a}{4m}\int d^{4}x\overline{\psi }_v\gamma ^{5}\sigma _{\alpha \beta } \psi_v \left( \theta F^{\alpha \beta
}\right) 
\nonumber \\
 && \rightarrow \frac{ieg_a}{2m}\int d^{4}x{\psi }^\dagger\sigma_{i}\psi 
\left( \theta E_{i}\right)
\eea
and:
\bea
& & \!\!\!\!\!\!\!\!\!\! -\frac{eg_a}{2m}\int d^{4}y\int d^{4}x\partial^\beta \overline{\psi }_{v} 
\left( \gamma _{5}\sigma _{\beta \rho }\right) \psi _{v}
 \cdot
 \nonumber \\
& &  \qquad \qquad \qquad G(x-y)\partial _{\gamma }[\theta(y)F^{\rho\gamma}(y)] 
\nonumber \\
&&  \rightarrow 
 \frac{ieg_a}{2m}\int d^{4}y\int d^{4}x\;\partial _{i}\left( \overline{%
\psi }^{\dagger }\sigma _{i} \psi (x)\right)
 \nonumber \\
& &  \qquad \qquad \qquad
G(x-y)\cdot
\left( \theta \partial _{j}E_{j}(y)\right) 
\eea
where the latter spinors (without subscript, ${}_v$) are two-component spinors. These 
expressions are
equivalent to eq.(\ref{finalPS}).
The result of eq.(\ref{georgirep}) is somewhat more general than
eq.(\ref{finalPS}), the latter refering to the frame with electron and axion both having 
$4$-velocity $v_\mu$.


\section{Anomalies, OEDM's, and Axion Decoupling}

\subsection{The General Form of the Action}

Presently we give a general argument as to why the nonlocal term
occurs, and we will derive the general structure of the action by
simply demanding the shift-symmetry of the axion.

The effect we are discussing arises perturbatively, from the
interaction of the axion anomaly with a magnetic moment 
and therefore it must share subtleties
 with the anomaly itself. 
The anomalous interaction
of the axion with electromagnetic fields
is written in eq.(\ref{zero}):
\beq
\label{zero1}
\frac{g_a}{2}\epsilon_{\mu\nu\rho\sigma}\int d^4 x\;\theta(x) 
F^{\mu\nu}{F}^{\rho\sigma}
\eeq
where  $\theta(x)=a(x)/f_a$ .  

We can perform a transformation  that shifts the axion as $\theta \rightarrow \theta+\phi $
where $\phi$ is an arbitrary angle. If $\phi$ is constant the theory is invariant
under this shift, since the anomaly only causes the action of eq.(\ref{zero1})
to shift by a total divergence, which has  no effect
in perturbation theory {
(a total divergence, or ``surface term,'' is $\propto$ the
total incoming momentum less the total outgoing momentum of the Feynman diagram). } Therefore, we conclude,
$a/f_a$ must be derivatively coupled in any perturbative process.

We see  this in
eq.(\ref{zero1}) if we integrate by parts and write:
\beq
\label{zero2}
-g_a\epsilon_{\mu\nu\rho\sigma}\int d^4x\;\partial^\mu \theta(t) 
A^\nu{F}^{\rho\sigma}
\eeq
where $A_\mu$ is the vector potential, and ${F}_{\mu\nu}=\partial_\mu A_\nu - \partial_\nu A_\mu$.
Eq.(\ref{zero2}) displays the derivative coupling
of the axion, however the manifest gauge invariance
of electromagnetism is lost. Alternatively, 
eq.(\ref{zero1}) maintains the gauge invariance of QED, but does
not display the $U(1)_{PQ}$ ``shift'' symmetry of the pNGB. 
Both symmetries are present in perturbation theory
since eq.(\ref{zero1}) and eq.(\ref{zero2}) differ only by
surface terms. 
We observed in Section II that
this feature is shared by the Feynman diagram
that generates a perturbatively induced
OEDM.

We generally define the covariant OEDM of the electron 
(or similarly for any
other object) {
in an arbitrary frame in which the cosmic axion field 
has four-velocity $u_\mu$, \ie, $\theta(x)=\theta_0\cos(u_\mu x^\mu)$,
as:
\beq
\label{zero3}
S' = g'\int d^4 x\; \theta(x) S_{\mu\nu}F^{\mu\nu}
\eeq
}
where  $S_{\mu\nu}$ is an antisymmetric odd parity dipole density
 ($S^0_{\mu\nu} \sim \bar{\psi}\sigma_{\mu\nu} \gamma^5\psi$;   
$S_{\mu\nu}$ will be defined in terms of $S^0_{\mu\nu}$ momentarily).

Since this result arises perturbatively from the anomaly, it
must have the shift symmetry in common with the anomaly. In particular
we must be able to  
move a derivative exclusively onto $\theta$ via integration by parts:
\beq
\label{zero3.2}
S' = -2g'\int d^4 x\; \partial^\mu\theta(x) S_{\mu\nu}A^\nu
\eeq
We see that the gauge field, $A^\mu$, now appears
explicitly,  exactly as happens in the
case of the anomaly itself.

However,  in order for  eq.(\ref{zero3.2}) to be valid,
a constraint must be satisfied:
\beq
\label{zero4}
\partial^\mu S_{\mu\nu}=0
\eeq
A  general solution to this constraint can be written as a nonlocal form:
\beq
\label{zero5}
S_{\mu\nu}=S^0_{\mu\nu}(x)+\int d^4 y \; \partial_{[\mu} G(x-y) \partial^\lambda S^0_{\nu]\lambda }(y)
\eeq
(note antisymmetrization in $\mu$ and $\nu$). Eq.(\ref{zero4}) satisfies 
$\partial^\mu S_{\mu\nu}=0$ for any antisymmetric $S^0_{\mu\nu}$, 
where 
the Green's function satisfies:
\beq
\partial^2 G(x-y) = \delta^4(x-y)
\eeq
and $G(x-y)$  is 
as defined in eq.(\ref{Green}).

{
The action with eq.(\ref{zero5}) thus becomes;
\bea
\label{zero6}
S'& = & g'\int d^4 x\;d^4 y\; \theta(x) F^{\mu\nu}(x) 
\nonumber \\
& & \!\!\!\!\!  \!\!\!\!\!   \times
\left[S^0_{\mu\nu}(x)\delta^4(x-y)+ \partial_{[\mu} G(x-y) \partial^\lambda S^0_{\nu]\lambda }(y) \right]
\eea
and  $S^0_{\mu\nu} = \bar{\psi}\sigma_{\mu\nu} \gamma^5\psi$ is local.
We can transpose the integrand, 
using $\partial_xG(x-y)=-\partial_yG(x-y)$ and
performing integrations by parts in $y$,  to obtain:
\bea
\label{zero7}
S' & = & g'\int d^4 x\; \theta(x)S^0_{\mu\nu}(x)F^{\mu\nu}
\nonumber \\
& & 
\!\!\!\!\!\!\!\!\!\!\!\!\!\!\!\!\!\!\!\!\!\!\!\!\!
+ 2\int d^4 x\int d^4 y \;  S^0_{\rho\nu}(x) G(x-y) 
\partial^\rho \partial^{[\mu}(\theta(y) F_{\mu}^{\;\;\nu] }(y))
\eea
This is seen to be equivalent, \eg, to the Georgi formalism result of eq.(\ref{georgirep})
after an integration by parts, where we have $g'=eg_a/4m$ and $S^0_{\mu\nu}=
\overline{\psi }_{v}
\gamma _{5}\sigma _{\mu\nu} \psi _{v}$. 
The role of the nonlocal term is therefore to maintain the shift symmetry
of the axion. 

The nonlocal term reduces further in certian limits.
In the limit of a constant $\theta(x)$  we  have 
$ \partial^{\mu}(\theta(y) F_{\mu}^{\;\;\nu }(y))=\theta\partial^{\mu} F_{\mu}^{\;\;\nu }(y)
=j^{\;\;\nu }(y)$, where $j^{\nu }(y)$ is a source current for the electromagnetic field.
Since $\partial^2 G = \delta $,  the vector potential is given by:
\bea
\label{zero7.1}
A_\mu(x)=\int d^4 y \;  G(x-y) j_{\mu }(y)
\eea
Hence, for constant $\theta$:
\beq
S'\rightarrow \int \theta S_{\mu\nu}F^{\mu\nu}-\int \theta S_{\mu\nu}F^{\mu\nu}=0
\eeq
which is the simplest statement of decoupling.

We can also consider  a  static limit
for the electron in which:
\bea
\int dy^0\; G(x-y) & = & \int \frac{d^3\vec{k}}{(2\pi)^3} \frac{e^{-i\vec{k}\cdot (\vec{x}-\vec{y})}}{\vec{k}^2 }
\nonumber \\
& = & \frac{1}{4\pi } \frac{1}{|\vec{x}-\vec{y}|}\equiv -\frac{1}{\vec{\nabla}^2}
\eea
where $\vec{\nabla}^2(1/R)=-4\pi \delta^3(R)$ (e.g, see  \cite{Jackson} section 6.4).
We find by explicit calculation
of Fig.(1) that $S^0_{\mu\nu}$ is the
a local electric dipole density operator,  $ \propto i\bar{\psi}\gamma^5 \sigma_{\mu\nu} \psi$).

For a purely spatially constant but time dependent axion field, $\theta(t)$ and
a  nonrelativistic, static
electric dipole moment,   $g'S^0_{0i}=-g'S^0_{i0}=\half {P}_i(x) $, 
where  $\partial_t {P}_i(x) =0 $ this becomes:
\beq
\small
S^{\prime }=g'\int d^{4}x\; \theta (t)\left( \overrightarrow{P}\cdot 
\overrightarrow{E}+\overrightarrow{\nabla }\cdot \overrightarrow{P}\left( 
\frac{1}{\overrightarrow{\nabla }^{2}}\right) \overrightarrow{\nabla }\cdot 
\overrightarrow{E}\right) 
\eeq
In an arbitrary gauge,
$\overrightarrow{E}=\overrightarrow{\nabla }\varphi
-\partial _{t}\overrightarrow{A}$ , we see that:
\bea
&& S^{\prime }  =g'\int d^{4}x\; \theta (t)\overrightarrow{\nabla }\cdot (
\overrightarrow{P}\varphi) 
\nonumber \\
& &\!\!\!\!\!\!\!\!  +g'\int d^{4}x\; \partial _{t}\theta
(t)\left( \overrightarrow{P}\cdot \overrightarrow{A}+\overrightarrow{
\nabla }\cdot \overrightarrow{P}\left( \frac{1}{\overrightarrow{\nabla }^{2}}
\right) \overrightarrow{\nabla }\cdot \overrightarrow{A}\right) 
\eea
Therefore, $S'$ becomes a total divergence in the
limit $ \partial _{t}\theta (t)\rightarrow 0.$  
In particular, the first term is a total divergence in space 
for a spatially constant $\theta(t)$.

In the case of a source-free  electric field, $\nabla\cdot \vec{E}=0$,
this becomes:
\bea
\label{xone}
g'\int d^{4}x\; \theta (t) \overrightarrow{P}\cdot 
\overrightarrow{E}
\eea
indistinguishable
from a simple electric dipole moment interaction.
Because $\overrightarrow{E}=-\partial_t\vec{A}$,
we can integrate this expression by parts to write:
\bea
\label{xtwo}
g'\int d^{4}x\; \partial_t\theta (t) \overrightarrow{P}\cdot 
\overrightarrow{A}
\eea
and manifest gauge invariance is lost, just as in the case of the anomaly
when the derivative is placed on the axion field.
Eqs.(\ref{xone}) and (\ref{xtwo}) differ only by surface terms
that are irrelevant to perturbation theory.

{
Note that if the electric field has a static source,
such as an electric charge (\eg, an atomic nucleus), located at $\vec{x}_0$,
$\overrightarrow{\nabla }\cdot 
\overrightarrow{E}(y)=Q\delta^3(\vec{y}-\vec{x})$
and, ignoring the explicit $\partial_t \theta\sim m_a$ terms in
the nonlocal component  of eq.(\ref{zero7}),  then quasistatic action reduces to:
\bea
\label{xone.1}
g'\int d^{4}x\; (\theta (t) -\theta (t-|\vec{x}-\vec{x}_0|)\overrightarrow{P}(x)\cdot 
\overrightarrow{E}(x) 
\eea
Here we see that the nonlocal term subtracts a retarded axion field from
the local value of the axion field.  This difference vanishes as $m_a\rightarrow 0$.
This is the way the decoupling is generally maintained in solutions to Maxwell's equations
in classical configurations, such as RF cavities, as we see in the next few sections.
This result, however, implies that the electron OEDM may be difficult to probe in atomic experiments
(such as the ACME experiment, \cite{ACME})  
where $|\vec{x}-\vec{x}_0|\sim r_{Bohr}$ is the Bohr radius, since  
$(\theta (t) -\theta (t-|\vec{x}-\vec{x}_0|)\sim m_a r_{Bohr} <<1$.

We emphasize that {\em a nonlocal operator structure in electrodynamics is not novel}. It is encountered 
in the ``transverse electromagnetic current'' in
QED, \eg, when we quantize in radiation gauge (see, eg, \cite{Jackson}, section 6.3 and eq.(6.28)).  
In that case
the nonlocal term is essential to maintain the causality of the theory in this gauge.
The transverse current occurs when we have Coulombic sources
and a nonzero, time dependent component of the vector potantial, $A_0$. 
$A_0$ has no time derivatives in the action and
is therefore an instantaneously propagating
field, and cannot represent a physical out-going on-shell photon. 
The equation
of motion for $A_0$ is $\vec{\nabla}^2 A_0 =-\rho(x)$, where $\rho(x)$ is a
charge density.  If we want to allow time dependent $A_0$, then 
$\nabla^2  \partial_0 A_0 =  -\partial_0 \rho(x,t)$, 
but from current
conservation we have $\partial_0 \rho=\nabla\cdot\vec{j}$ 
where $\vec{j}$ is the 3-current.
 Hence, we have  $\partial_0 A_0 = -(1/\vec{\nabla}^2 )\vec{\nabla}\cdot \vec{j} $.
 This means that if $A_0$ is to be time dependent, then there must
necessarily be a 3-current, hence a vector potential, $\vec{A}$. 
We impose
the condition $\vec{\nabla}\cdot\vec{A}=0$. $\vec{A}$ satisfies $ (\partial_0^2-\nabla^2) \vec{A} 
-\vec{\nabla}\partial_0 A_0= \vec{j}$ (the equation of motion
of $A_0$ is unmodified by this). This is often written as 
$ (\partial_0^2-\nabla^2) \vec{A} 
 = \vec{j}_T $.
where $\vec{j}_T $ is the ``transverse current'' \cite{Jackson} which takes the form 
$\vec{j}_T = \vec{j}-\vec{\nabla}(1/\nabla^2) \vec{\nabla}\cdot \vec{j} $.
$\vec{j}_T$ is identically conserved with the nonlocal term, is consistent
with the radiation gauge condition  $\vec{\nabla}\cdot\vec{A}=0$, and thus
gauge invariance is maintained.  From this, Lorentz invariance
is also maintained.

Thus, introducing $A_0$ time dependence requires a nonlocal
correction to the current to maintain a conserved current, hence gauge invariance.
This is formally similar to the structure seen in eq.(\ref{zero5}) which
maintains the shift symmetry (analogue of gauge symmetry) of the axion.  
 The apparent nonlocality is arising because we are
treating the ``vacuum'' as effectively containing a space-time dependence,
through the background cosmic classical $\theta(t)$ field.  
Physical amplitudes thus inherit a nonlocal dependence upon the history of the vacuum.}

\subsection{
Axion in an Infinite Volume Static Magnetic Field and Inherent Nonlocality}

{
The ``integral form of axion decoupling,'' as we have seen above
arising from the nonlocal term, is a general feature
of the solutions to the Maxwell equations in various practical
situations. }
Suppose we have an infinite universe which contains
a uniform static magnetic field $\overrightarrow{B}_{0}$ and a background
oscillating 
classical axion field, $a(t)/f\equiv \theta \left( t\right) =\theta _{0}\cos \left(
m_{a}t\right)$ (this can be considered as an infinite volume
limit of an RF cavity experiment as we do below).
We consider the sources for $\overrightarrow{B}_{0}$
to be far away from the region of interest and therefore
$ \overrightarrow{\nabla}\times \overrightarrow{B}_{0}=0$.
The decoupling, $m_a\rightarrow 0$ limit,
becomes somewhat subtle, even in this case. 
We can analyze this classically.  

The axion anomaly will generate an electromagnetic field
of the form:  $\overrightarrow{E} = \overrightarrow{E}_{r} $
and $\overrightarrow{B} = \overrightarrow{B}_{0}+\overrightarrow{B}_{r} $
where $\overrightarrow{E}_{r}$ and $\overrightarrow{B}_{r}$   are oscillating 
``response fields.'' 
Maxwell's Equations in these fields
become: 

\noindent
Maxwell (1):
\noindent \beq \label{max1} \qquad \overrightarrow{\nabla }\times \overrightarrow{B}%
_{r}-\partial _{t}\overrightarrow{E}_{r}=-g_a\overrightarrow{%
B}_{0}\left( \partial _{t}\theta \right) \eeq

\noindent
Maxwell (2) 
\noindent \beq  \label{max2}  \qquad \overrightarrow{\nabla }\times \overrightarrow{E}%
_{r}+\partial _{t}\overrightarrow{B}_{r}=0\eeq

\vskip 0.25cm
\noindent
and $\ \ \overrightarrow{\nabla }\cdot \overrightarrow{B}_{r}=%
\overrightarrow{\nabla }\cdot \overrightarrow{E}_{r}=0.$

The Maxwell equations are coupled first order inhomogeneous
differential equations.  They are consistent 
with the decoupling of the axion as $m_a\rightarrow 0$, since the source term is proportional
to $\partial _{t}\theta$.
The vector potential in Coulomb gauge likewise
satisfies:  $\partial _{t}^{2}\overrightarrow{A}_{r}-\overrightarrow{%
\nabla }^{2}\overrightarrow{A}_{r}=-g_a\overrightarrow{B}%
_{0}\left( \partial _{t}\theta \right) $ \ where \ $\overrightarrow{E}%
_{r}=-\partial _{t}\overrightarrow{A}_{r}$ and $\nabla\cdot\overrightarrow{A}_{r}
=0$, a single second order inhomogeneous differential equation.
For an infinite universe filled with the magnetic field
$\overrightarrow{B}_{0}$ we have translational invariance in space.

The Maxwell equations have a particular solution that is consistent
with the symmetry of spatial translational invariance:
\bea
\overrightarrow{E}_{r} & = & g_a\overrightarrow{B}_{0}\int_{0}^{t}d\tau 
\partial _{\tau }\theta (\tau )=g_a
\overrightarrow{B}_{0}\left( \theta (t)-\theta (t_0)\right) \nonumber \\
\overrightarrow{B}_{r} & = & 0
\eea
The solutions necessarily require the specification
of a single boundary condition at an inital time,
which we take to be  $\overrightarrow{E}_{r}(t_0)=0$ 
and  $\overrightarrow{B}_{r}(t_0)=0$ .
We emphasize that this is a non-propagating solution (since $\overrightarrow{B}_{r}  =  0$) and represents a time
dependent ``dual rotation'' of $\overrightarrow{B}_{0}\rightarrow \overrightarrow{E}_{r}$
\cite{Hill1}.

 The dependence upon the initial condition
introduces an apparent nonlocality, or ``history'', into the observed $\overrightarrow{E}_{r}(t)$.
Clearly $\overrightarrow{\nabla }\times \overrightarrow{E}_{r}=0$ hence $B_r=0$.
Note that the  vector potential is 
nonlocal,  satisfying
\beq
\overrightarrow{A}_{r}=g_a\overrightarrow{B}%
_{0}\int_{0}^{t}d\tau ^{\prime }\int_{0}^{\tau ^{\prime }}d\tau \partial
_{\tau }\theta (\tau )
\eeq
Our infinite universe with a static magnetic field and an oscillating axion
field has acquired an oscillating electric field.  This is an example of
a general ``theorem:'' {\em Any magnetic moment becomes an oscillating electric moment
in the presence of the oscillating axion.  }
The magnetic field need not be
restricted to a constant all-space filling form as in this toy universe example;
it can be, \eg, the local field surrounding a magnetic dipole moment, as we
can see classically.  The effect of the axion is to produce a time dependent electric dipole
for the electron.

\subsection{Axion Induced Electric Field in an RF Cavity}

To detect a cosmic axion signal we can  deploy  a very large 
constant, externally applied magnetic field $\overrightarrow{B}_{0}$ within
a resonant cavity.  The solutions to the Maxwell 
equations for the response fields will always involve the same  particular solution
we just encountered in the toy universe, but also now 
includes homogeneous solutions that are required to implement the boundary conditions
of the cavity.

Maxwell's equations for the reponse fields are as 
in eqs.(\ref{max1},\ref{max2}).
We now have conducting boundary conditions at the cavity wall, $r=R$:
$\overrightarrow{E}(r=R)=0 $  (and since the form of the $B$-field is parallel
to the cavity axis, we have no constraint upon $\overrightarrow{B}(r=R)$).

With cylindrical coordinates, ($\rho ,\phi ,z$ ), we find
a cylindrically symmetric solution:
\bea
\overrightarrow{E} & =& \left( kJ_{0}(\rho m_{a})+g_a\right) B_{0}%
\widetilde{\theta }\widehat{z}
\nonumber \\
 \overrightarrow{B} & =& kJ_{1}(\rho m_{a})B_{0} \frac{\partial
_{t}\widetilde{\theta }}{m_{a}}\ \widehat{\phi }
\eea
The electric field has the form of our free space  solution proportional to $g_{A }$, plus a
homogeneous cavity mode solution proportional to $k.$

We now apply conducting boundary conditions at the cavity wall,
$\ \overrightarrow{E}(\rho =R)=0$. We thus determine $k:$
\bea
\overrightarrow{E}=-g_a\left( \frac{J_{0}(\rho m_{a})}{
J_{0}(Rm_{a})}-1\right) B_{0}\widehat{z}\widetilde{\theta } 
\nonumber \\
 \overrightarrow{B}=-g_a\frac{%
J_{1}(\rho m_{a})}{J_{0}(Rm_{a})}B_{0}\widehat{\phi }\left( \frac{%
\partial_{t}\widetilde{\theta }}{m_{a}}\right) 
\eea
Note again that the solution vanishes as  $m_{a}\rightarrow 0$ since we
define $\widetilde{\theta }(t)=\int_{t_0}^{t}d\tau \partial _{\tau
}\theta (\tau ),$ where $t_0$ is an earlier
time at which $ \overrightarrow{E}(t_0)= 0 $. The solutions is 
therefore intrinsically nonlocal in time.
Henceforth we will omit the understood tilde on $\widetilde{\theta }$,
\ie, $\theta \rightarrow  \theta(t) - \theta(t_0)$.

This is an idealized solution with a perfect resonant behavior, i.e., for a
special cavity
satisfying  $J_{0}(Rm_{a})=0$  the solution has an apparent infinite amplitude. 
Of course, in reality
the amplitude is damped by dissipation.
This results in a  finite $Q$ value,
and  modifies the solution to:
\bea
\label{solutionQ}
\overrightarrow{E}
& = &  -g_a\left( \frac{J_{0}(\rho m_{a})}{F}-1\right) B_{0}%
\widehat{z}{\theta }
\nonumber \\
\overrightarrow{B} & = & 
-g_a\frac{
J_{1}(\rho m_{a})}{F}B_{0}\widehat{\phi }\left( \frac{\partial _{t}
{\theta }}{m_{a}}\right)
\eea
where 
\beq
F=\sqrt{\left( J_{0}(Rm_{a})\right) ^{2}+1/Q^{2}}
\eeq
{ 
Note that the total oscillating field energy in the cavity is $\propto Q^2$.
The finite $Q$ arises if we include a resistive damping
term in the Maxwell equations. 
In an idealized perfect cavity the oscillating fields are the electric field    $90^o$ out
of phase from the magnetic field, and hence a vanishing time averaged Poynting vector.
There is, however, with finite $Q$ there is an induced, small ${\cal{O}}(1/Q)$, temporal
phase shift, that
we have not written, such that
$\overrightarrow{E}$ and $\overrightarrow{B}$ are not exactly $90^o$
out of phase. This allows the 
Poynting vector at the walls of the cavity to average to a 
nonzero result. Hence  power is extracable at a rate $\propto Q$. 
If we attempt to extract power at  faster rate then $Q$ will decrease
since the dominant 
power loss mechanism becomes the radiative extraction itself. 
We will consider some quantitative aspects of the RF cavity 
solution in Section V.
}

\section{Electric Dipole Radiation from a Stationary Electron}

\subsection{Classical Calculation}

First we consider the electric dipole
radiation from a classical magnetic moment immersed in
the axion field.  This calculation has  validity for intense 
classical magnetic sources.
It cannot be adapted to
the case of an electron which
requires the quantum calculation.  Nonetheless,
it is instructive to compute
and compare it with the quantum case in Section IV.B.
We will see that the radiation
is generated by the physical OEDM. The magnetic dipole
field surrounding the source, though it appears
on the rhs of Maxwell's equations as a source term in the axion background,  does not itself radiate.
Instead, this is associated with the non-radiating particular solution encountered 
in the previous section, and it allows implementation of various
boundary conditions at short distance for the electric field.

The  standard Maxwell equations with axion anomaly source
in the presence of an arbitrary static, local magnetic field $\overrightarrow{B}_{0}(\vec{r})$
are:
\vskip 0.25cm

\noindent
Maxwell (1):
\noindent 
\beq \label{maxx1} \qquad \overrightarrow{\nabla }\times \overrightarrow{B}%
_{r}(\vec{r},t)-\partial _{t}\overrightarrow{E}_{r}(\vec{r},t)=-g_a\overrightarrow{%
B}_{0}(\vec{r})\partial _{t}\theta(t)  \eeq

\noindent
Maxwell (2) 
\noindent 
\beq  \label{maxx2}  \qquad \overrightarrow{\nabla }\times \overrightarrow{E}%
_{r}(\vec{r},t)+\partial _{t}\overrightarrow{B}_{r}(\vec{r},t)=0
\eeq
and $\overrightarrow{\nabla }\cdot \overrightarrow{B}_{r}=%
\overrightarrow{\nabla }\cdot \overrightarrow{E}_{r}=0.$

Presently $\overrightarrow{B}_{0}(\vec{r})\equiv \overrightarrow{\nabla }\times \overrightarrow{A
}_{0}(\vec{r})$ will be  the static field  of a 
classical solenoidal  magnet, centered at the origin, $\vec{r}=0$,
where $ \overrightarrow{A
}_{0}$  can be written formally in terms of  a magnetic dipole source $\overrightarrow{m}_{0}=
\overrightarrow{m}\delta ^{3}(\overrightarrow{r}) $, (see ref.\cite{Jackson}, Chapter 5.6).
We have formally:
\beq
\label{source1}
-\overrightarrow{\nabla }\times \overrightarrow{B%
}_{0}=\overrightarrow{\nabla }^{2}\overrightarrow{A}_0=\overrightarrow{\nabla }%
\times \overrightarrow{m}_{0}
\eeq
hence:
\beq
\overrightarrow{A}_0 = \frac{1}{4\pi} \frac{\overrightarrow{m}\times 
\overrightarrow{r} }{r^3}
\eeq
and:
\beq
\label{source3}
\overrightarrow{B}_0= - \frac{1}{4\pi}\left[\frac{8\pi}{3}\overrightarrow{m}\delta^3(\vec{r})+ \left( \frac{1}{r^3} \right) \left( 
\overrightarrow{m}-\frac{3\overrightarrow{r}\left( \overrightarrow{r}\cdot 
\overrightarrow{m}\right) }{r^2}\right) \right]
\eeq
(see the magnetostatics
discussion in Jackson, eqs.(5.55-5.64) \cite{Jackson};
Jackson states that this is a purely classical construction and
cannot be unambiguously applied to a quantum mechanical electron).

{
A subtlety arises in the present case 
with the particular solution encountered in Section III.
Consider a ``sourceless dipole field,'' \ie, one
in which there is no $\overrightarrow{m}\delta^3(\vec{r}) $ term:
\beq
\widehat{B}_0= - \frac{1}{4\pi} \left( \frac{1}{r^3} \right) \left( 
\overrightarrow{m}-\frac{3\overrightarrow{r}\left( \overrightarrow{r}\cdot 
\overrightarrow{m}\right) }{r^2}\right)
\eeq
We consider Maxwell's equations,  replacing  $\overrightarrow{B}_0$ by $\widehat{B}_0$ on the rhs of eq.(\ref{maxx1}).
We then have the particular solution to the vacuum Maxwell equations:
\bea
\label{Er0}
\ \overrightarrow{E}_{r}(\vec{r},t)&=&-g_a\widehat{B}_0(\vec{r})\theta(t)
\eea
and:
\bea
\label{Br0}
\overrightarrow{B}_{r}(\vec{r} ,t)=0
\eea 
 This is a solution since:
\beq
0= \overrightarrow{\nabla }\times\left( \frac{%
\overrightarrow{m}}{r^{3}}-3\frac{\overrightarrow{r}\left( 
\overrightarrow{m}\cdot \; \overrightarrow{r}\right) }{r^{5}}\right) 
\eeq
(note that any potential
severe surface term singularity arising here, \eg, such as in $\overrightarrow{\nabla }_i(r_j/r^5) $,
will be $\propto \delta_{ij}$ or  $\propto r_{i}r_{j}$, but contracted with $ \epsilon_{ijk}$ 
from the cross-product, and hence zero). 
However, since $\overrightarrow{B}_{r}(\vec{r} ,t)=0 $,   this is a nonpropagating solution. It exists
for any background static magnetic field satisfying $\overrightarrow{\nabla }\times \overrightarrow{B%
}_{0}=0$ .  We will require incorporating this non-propagating
solution into our full radiation solution momentarily.

The existence of this
solution has an important implication: The  sourceless dipole field surrounding the 
origin, {\em does not lead to radiation}. The radiation comes
only when we have a nonzero magnetic field $\overrightarrow{%
B}_{0} $ which curls around a nonzero  dipole source term.  This is analogous to 
the infinite universe with a constant magnetic field versus the RF cavity:
it is the conducting wall of the cavity that enforces
the boundary condition that produces the  resonant magnetic and
 electric fields. 
}

We can now solve the Maxwell equations using retarded Green's functions
for a vector potential, $A_r(\vec{r},t)$, describing the 
radiative oscillating response fields in Coulomb gauge.
The analysis mostly follows the textbook derivations as
in \cite{Jackson}, Chapter 9, but involves the non-propagating
solution described above. 
In what follows, we will pass to complex notation 
where $\theta(t)=\theta_0\exp(im_at)$, and the physical response fields will
be the real parts of the complex $ \overrightarrow{E}_{r} $ and $ \overrightarrow{B}_{r}$. 
The radiated electromagnetic fields $\overrightarrow{B}%
_{r}(\vec{r},t)$ and $\overrightarrow{E}%
_{r}(\vec{r},t)$   are obtained as follows:
\bea
\label{Er}
\ \overrightarrow{E}_{r}(\vec{r},t)& = &
-\frac{1}{4\pi}g_a\theta (t)\exp (-im_{a}|\overrightarrow{r}|)\nonumber \\
& &  \!\!\!\!\!  \!\!\!\!\!  \!\!\!\!\!  \!\!\!\!\!  
\!\!\!\!\!  \!\!\!\!\!  \!\!\!\!\!  \!\!\!\!\! 
\cdot \left( 
\left[ 1-\exp (im_{a}|\overrightarrow{r}|) +im_{a}r\right] \left( \frac{%
\overrightarrow{m}}{r^{3}}-3\frac{\overrightarrow{r}\left( 
\overrightarrow{m}\cdot \; \overrightarrow{r}\right) }{r^{5}}\right)\right. \nonumber \\
& &\left.\qquad 
-m_{a}^{2}\left( \frac{\overrightarrow{m}}{r}-
\frac{\overrightarrow{r}(\overrightarrow{m}\cdot \overrightarrow{r})}{r^3}\right) \right)
\eea
and:
\bea
\label{Br}
\overrightarrow{B}_{r}(\vec{r} ,t)=\frac{1}{4\pi}g_a\partial
_{t}\theta (t)\exp (-im_{a}|\overrightarrow{r}|)
\nonumber \\
& &  \!\!\!\!\!  \!\!\!\!\!  \!\!\!\!\!  \!\!\!\!\!  \!\!\!\!\!  \!\!\!\!\!  \!\!\!\!\!  \!\!\!\!\!  
\cdot \;
\overrightarrow{m} \times
\left( \frac{\overrightarrow{r}}{r^{3}}+\frac{im_{a}\overrightarrow{r}}{r^{2}%
}\right) 
\nonumber \\
\eea
One can readily verify that eqs.(\ref{Er}, \ref{Br}) satisfy the Maxwell
equations eqs.(\ref{maxx1}, \ref{maxx2})  with the source term $ -g_a\overrightarrow{%
B}_{0}(\vec{r})\left( \partial _{t}\theta(t) \right) $.  
Notice that the second term in the brackets $\left[...\right]$ in eq.(\ref{Er})
is the non-propagating  solution of eq.(\ref{Er0})

{
Taking the near-zone limit yields:
\bea
\label{Ernear}
\overrightarrow{E}_{r}(\vec{r},t) & \rightarrow &
0
\eea
which vanishes due to cancellation
with the particular solution, and:
\beq
\label{Brnear}
\overrightarrow{B}_{r}(\vec{r},t)\rightarrow \frac{1}{4\pi}g_aim_a\theta
(t)\left( \ \overrightarrow{m}\ \times \frac{\overrightarrow{r}}{r^{3}}%
\right)
\eeq
From the 
near-zone limit we can confirm the Maxwell equations
and read-off the source structure: 
\bea
\nabla\times\overrightarrow{B}_{r}- \partial_t\overrightarrow{E}_{r}& = & -\frac{1}{4\pi}g_{a}\partial _{t}\theta (t)
\left(-\frac{8\pi}{3}\overrightarrow{m}\delta^3(\overrightarrow{r})\right.
\nonumber \\
& &\qquad \left.
+ \frac{\overrightarrow{m}}{r^{3}}-3\frac{%
\overrightarrow{r}\text{ }\left( \overrightarrow{m}\cdot \overrightarrow{r}%
\right) }{r^{5}}\right)  
\eea
Thus we see that the propagating radiation is due to the physical OEDM
source, which induces the curling magnetic field, 
and is not due to the dipole magnetic field surrounding the source. 
}

In the far-zone we have:
\bea
\overrightarrow{E}_{r}(\vec{r},t)\rightarrow \frac{g_a m_{a}^{2}}{4\pi}\theta (t)\exp
(-im_{a}|\overrightarrow{r}|)\left( \frac{\overrightarrow{m}}{r}-\frac{%
\overrightarrow{r}}{r^{2}}\frac{\overrightarrow{m}\cdot \overrightarrow{r}}{r%
}\right) 
\nonumber \\
\eea
\bea
\overrightarrow{B}_{r}(\vec{r},t)\rightarrow -\frac{g_a m_{a}^{2}}{4\pi}\theta (t)\exp
(-im_{a}|\overrightarrow{r}|)\left( \ \overrightarrow{m}\ \times \left( 
\frac{\overrightarrow{r}}{r^{2}}\right) \right) 
\nonumber \\
\eea

These 
are seen to be formally equivalent
to the {\em electric dipole radiation fields },
where our magnetic moment $\vec{m}$ replaces the electric moment $\vec{p}$
in Jackson, \cite{Jackson} eq.(9.18).  Hence, the source of the radiation is
the axion induced  OEDM.

From this we can compute the cycle averaged Poynting vector, $\overrightarrow{K}
=<\overrightarrow{E}_{r}\times \overrightarrow{B}_{r}> $:
\bea
\overrightarrow{K} =\frac{1}{32\pi^2}g_a^{2}m_{a}^{4}\theta _{0}^{2}\left[\left( 
\frac{\overrightarrow{r}}{r^{2}}\right) \left( \frac{\overrightarrow{m}^{2}}{%
r}-\frac{\left( \overrightarrow{m}\cdot \overrightarrow{r}\right) ^{2}}{r^{3}%
}\right) \ \right] 
\eea
Using $\overrightarrow{m}=\mu_{Bohr}\vec{S}$,
the angular differential  emitted power, $P$, for
our classical electron is therefore given by the classical
dipole pattern (\cite{Jackson}, eq.(9.23)):
\bea
\label{power0}
\frac{dP}{d\Omega} = \frac{1}{32\pi^2}g_a^{2}m_{a}^{4}\theta
_{0}^{2}\mu _{\text{Bohr}}^{2}\sin^2\theta
\eea
The total the emitted power is then:
\bea
\label{power}
P_{tot}= \frac{1}{12\pi }g_a^{2}m_{a}^{4}\theta
_{0}^{2}\mu _{\text{Bohr}}^{2}.
\eea

\newpage
\subsection{Quantum Calculation}

The  quantum calculation is straightforward, but
as one would expect is also somewhat tricky.
  We will summarize
it presently.   From the Pauli-Schroedinger result we have an effective
action for radiation given by eq.(\ref{finalPS}):
\bea
\label{finalPS2}
&& =\frac{ieg_a}{2m}\int d^{4}x\; \psi ^{\dagger }\sigma _{k}\psi \; \theta (t)E_{k}(x)
\eea
where we can neglect the nonlocal term since 
$\partial _{i}E_{i}=0$. 

 The coherent axion field
$\theta(t) = \frac{\theta_0}{2}(e^{im_at}+e^{-im_at})$ has both incoming, $e^{-im_at}$
and outgoing $e^{+im_at}$ components, and we  drop the outgoing 
component for the process of conversion of an axion into a photon.
Likewise, we replace $E_{i}(x)=\partial_0A_i(x)$ by an outgoing photon $ik_0\epsilon_i e^{+ik_0t-i\vec{k}\cdot \vec{x}}$
with polarization $\vec{\epsilon}$ where $\vec{\epsilon}\cdot \vec{k}=0$. The electron is assumed to be
localized in
space and we write $\psi^{\dagger }\sigma _{k}\psi = \chi_f^{\dagger }\sigma _{k}\chi_i\delta^3(\vec{x})$
for initial and final two-component spinors $\chi_i$ and $\chi_f$.  

Consider an emission amplitude of a photon in the $xz$ plane
(which corresponds to the polar azimuthal angle, $\phi = 0$; we'll integrate over $\phi$ subsequently).
The photon  3-momentum is $\vec{k} = (\sin(\theta), 0 , \cos(\theta))$.
The $\delta^3(\vec{x})$
distribution implies that the 3-momentum of the photon is arbitrary, constrained only by the on-shell
condition and, finally, by energy conservation $k_0=|\vec{k}|=m_a$.
The photon can, in principle,  have two independent polarizations,
satisfying $\vec{\epsilon}_i\cdot\vec{k}=0$, which we can take to be
$\vec{\epsilon}_1=(-\cos(\theta), 0, \sin(\theta)) $ or  $\vec{\epsilon}_2=(0, 1, 0) $.

Consider first the case of the initial electron with spin up transitioning
to a final electron, also with spin up, hence $\chi_i=\chi_f=\left({\begin{array}{c} 1 \\ 0 \end{array}}\right)$.  
Then, we see that:  
\beq
\vec{\epsilon}_1\cdot\chi_f^{\dagger }\vec{\sigma}\chi_i=\sin(\theta)\;\;\;
\makebox{and} \;\;\; \vec{\epsilon}_2\cdot\chi_f^{\dagger }\vec{\sigma}\chi_i=0
\eeq
Therefore, only a photon of polarization $\vec{\epsilon}_1$ is emitted in this case.
So the amplitude for spin-up to spin-up is therefore:
\beq
A_{\uparrow  \uparrow }=\frac{1}{2}g_a\mu_{Bohr}\theta_0 k_0 \sin(\theta)\times 2\pi\delta(m_a-k_0)
\eeq
The corresponding transition rate is:
\bea
\Gamma_{\uparrow \uparrow } & = & \frac{1}{4}(g_a\mu_{Bohr}\theta_0)^2
\int
\frac{k_0^2\sin^2(\theta)d^3k}{(2\pi)^3 2k_0} (2\pi\delta(m_a-k_0))
\nonumber \\
& = & \frac{1}{4}(g_a\mu_{Bohr}m_a\theta_0)^2\int \frac{m_a}{2(2\pi)^2 } \sin^3(\theta)d\theta d\phi
\nonumber \\
& = & \frac{m_a}{12\pi}(g_a\mu_{Bohr}m_a\theta_0)^2
\eea
The emitted power is therefore: 
\beq
P_{\uparrow \uparrow }={m_a}\Gamma_{\uparrow \uparrow}=\frac{1}{12\pi}(g_a\mu_{Bohr}m^2_a\theta_0)^2
\eeq
This is identical to the classical case of eq.(\ref{power}).

However, there is, for a free electron, the possibility of
a spin flip.  Consider the case of the initial electron with spin up transitioning
to a final electron, with spin down, hence $\chi_i=\left({\begin{array}{c} 1 \\ 0 \end{array}}\right)$, 
$\chi_f=\left({\begin{array}{c} 0 \\ 1 \end{array}}\right) $
Now we see that  
\beq
\vec{\epsilon}_1\cdot\chi_f^{\dagger }\vec{\sigma}\chi_i=0\;\;\;\makebox{and}
\;\;\;\vec{\epsilon}_2\cdot\chi_f^{\dagger }\vec{\sigma}\chi_i=i
\eeq
Therefore, only a photon of polarization $\vec{\epsilon}_2$ is now emitted.
So the amplitude for spin-up to spin-down is therefore:
\beq
A_{\uparrow \downarrow}=\frac{i}{2}g_a\mu_{Bohr}\theta_0 k_0 2\pi\delta(m_a-k_0)
\eeq
The corresponding rate is:
\bea
\Gamma_{\uparrow \downarrow} & = &  \frac{1}{4}(g_a\mu_{Bohr}m_a\theta_0)^2\int \frac{m_a}{2(2\pi)^2 } \sin(\theta)d\theta d\phi
\nonumber \\
& = & \frac{m_a}{16\pi}(g_a\mu_{Bohr}m_a\theta_0)^2
\eea
The emitted power is therefore: 
\beq
P_{\uparrow \downarrow}={m_a}\Gamma_{\uparrow \downarrow}=\frac{1}{16\pi}(g_a\mu_{Bohr}m^2_a\theta_0)^2
\eeq
Hence, in free space an electron
will radiate with a total power given by
\beq
P_{total}=P_{\uparrow \uparrow}+P_{\uparrow \downarrow}=\frac{7}{48\pi} (g_a\mu_{Bohr}m^2_a\theta_0)^2
\eeq
For an electron constrained to remain spin-up the power is that of eq.(\ref{power}).
 
In the following we will be interested in polarized electrons, such as electrons in
ferromagnets.  Polarized electrons are generally sitting in a polarizing
$B$-field, and there is therefore an energy cost in flipping the spin.
Hence, we are justified in dropping the $P_{\uparrow \downarrow}$ rate in estimates
of aggregated electrons in magnetic materials emitting dipole radiation. 
The quantum calculation, where we restrict the final state to spin-up
for an initial spin-up state, gives identically the same
power rate as the classical calculation. 

It is interesting that the axion field can also cause electrons to absorb photons.
We will not further consider this ``axion induced cooling'' effect, 
but perhaps it, too, has some interesting  consequences.

\newpage
\section{Some Quantitive Estimates  
}

Presently we give some quantitative estimates
for possible detection schemes.  We will briefly
review the
RF cavity, but then focus exclusively on the possible
detection of the radiation from coherent
assemblages of magnets or large scale magnetic fields.  
We  plan a more detailed treatment  elsewhere \cite{Hill3}.

We begin by defining a useful
scaling parameter: $ \eta= (f_a/10^{12}\makebox{GeV}) $.
Axion parameters and conversions to natural units are as follows:

\vspace{0.2 in}
\noindent 
\begin{tabular}{||l|c|l||}
\hline
decay constant   & $f_a $   &  $ \;  \eta= (f_a/10^{12}\makebox{GeV}) $ \\ \hline
axion mass          & $\;\;m_a$                & $\; 6.02\times 10^{-15} \;\eta^{-1} $ GeV \\ \hline
axion wavelength $\lambda_a$ &  $\frac{2\pi \hbar}{m_ac}$  &   $20.6\; \eta $ cm \\ \hline
axion frequency $\nu_a$ &  $c/\lambda_A$  &   $1.46\times 10^9$ \; Hz\; \\ \hline
cosmic amplitude  &  $\;\;\theta_0\;\;$  &   $3.68\times 10^{-19}{\rho}_g$  \\ \hline
anomaly coefficient  &  $\;\; g_a\;\;$  &   $-2.26\times 10^{-3} $ assumed  \\ \hline
 magneton $\mu_{Bohr\;CGS}$ & $\;\frac{e\hbar}{2m_ec}\;$  &   $83.591$ GeV$^{-1}$  \\ \hline
$1$ statvolt/cm (cgs) &  & $ 6.92\times 10^{-20} $ GeV$^2$   \\ \hline
$1$ watt  &  &   $4.09\times10^{-15} $ GeV$^2$  \\ \hline
$1$ esla & & $ 6.92\times 10^{-16} $ GeV$^2$ \\ \hline
\end{tabular}

\vspace{0.05in}
{\begin{center}
{Table I.  Axion parameters and  conversions.\footnote{We assume 
the galactic halo energy density ${\rho}_g=\rho_{galaxy}/(0.3\;\makebox{GeV}/\makebox{cm}^3)$ ;
the cosmic axion field amplitude, $a(t)/f_a = \theta_0\cos(m_A t+\phi)$, where $\phi$ is arbitrary.
Note that $\theta_0$ is independent of $f_a$ \cite{Graham}.}
\footnote{
One must be careful, as usual, in the definition of the Bohr magneton.
In quantum electrodynamics we typically use
the SI system of units, in which  $e_{SI}^2/4\pi=\alpha=1/137$,
and field energy density [Poynting vector] is $(\overrightarrow{E}^2+\overrightarrow{B}^2)/2$
[$\overrightarrow{E}\times \overrightarrow{B}$].
Note that we define the  Bohr magneton  $\mu_{Bohr\;SI}=e_{SI}\hbar/2m_ec$ in the SI system (see eq.\ref{magmom}). 
If we choose CGS units where $e_{CGS}^2=\alpha$ the energy density 
[Poynting vector] is $(\overrightarrow{E}^2+\overrightarrow{B}^2)/8\pi$
[$\overrightarrow{E}\times \overrightarrow{B}/4\pi$] and the Bohr magneton is now 
$\mu_{Bohr\;CGS}=e_{CGS}\hbar/2m_ec$.
The definitions in SI vs MKS differ by a familiar factor of $\sqrt{4\pi}$,
\ie, $\mu_{Bohr\; SI}= \sqrt{4\pi}\mu_{Bohr\; CGS}$.
Put another way, our above SI calculation of radiated power, eq.(\ref{power}), 
yielded $P=(12\pi)^{-1}(\mu_{Bohr\;SI} ...)^2=(12\pi)^{-1}(\sqrt{4\pi}\mu_{Bohr\;CGS} ...)^2$. 
Had we used CGS field normalizations and computed the Poynting vector directly
we would have directly obtained the formula $P=(3)^{-1}(\mu_{Bohr\;CGS} ...)^2$.  
}
 }\end{center}}
\vspace{0.05in}

\noindent

\subsection{RF Cavity Energetics}

Let us  estimate the signal power of a resonant
 RF cavity experiment.  [This follows an estimate 
of Aaron Chou \cite{Chou2}
pertaining to the ADMX experiment]. 
The ADMX microwave cavity is cylindrical
and has a length of $L\sim 10^{2}$ cm, 
and radius of the cavity bore of $R\sim 25$ cm, therefore 
a cross sectional area of $\pi \left( 25\right) ^{2}=\allowbreak 1.96\times
10^{3}$ cm$^{2}$.
The volume of the cavity bore
is $V=\pi R^{2}L$ $\sim 1.96\times 10^{5}cm^{3}$.
We will assume $Q\sim 10^{5}$ , 
and an applied 
constant external magnetic field $B_{0}=7$ Tesla.  

We see from eq.(\ref{solutionQ}) that 
the oscillating
signal fields in an RF cavity are of order: 
\beq
|\overrightarrow{B}_{r}|\sim |\overrightarrow{E}_{r}|\sim g_a B_{0}\theta
_{0}Q \approx  5.8\times 10^{-12}\left(\frac{Q}{10^5}\right) \; \makebox{(cgs)}
\eeq
The total signal energy
in the cavity is therefore: 
\beq
{\cal{E}}_0\sim \frac{1}{8\pi }\left( g_aB_{0}\theta _{0}\right) ^{2}\times Q^{2}VC
\sim  1.85\times 10^{-19}\left(
Q/10^{5}\right) ^{2} \; \makebox{ergs}
\eeq
where $C$ is a form factor parameterizing
the shapes of the  cavity modes which
we take to be $C\approx 0.7$.

A damped driven simple harmonic oscillator
of natural frequency $\omega _{0}$, driving force $F_{0}\exp \left(
i\omega t\right) ,$ satisfies,
\beq
\partial _{t}^{2}\phi +\Gamma \partial _{t}\phi +\omega
_{0}^{2}\phi =F_{0}\exp \left( i\omega t\right) 
\eeq
hence: 
\beq
\phi={Re}(\phi _{0}\exp (i\omega t));\qquad \ \phi _{0}=\frac{
F_{0}}{\left( \omega _{0}^{2}-\omega ^{2}\right) +i\omega \Gamma }\
\eeq
On resonance this will have an energy stored in a cycle
of ${\cal{ E}}_{0}=|\phi _{0}\omega _{0}|^{2}/2$,
and energy lost in a cycle
(which must
be replaced in a steady state by the driving term), of ${\cal{ E}}_{lost}=$\ $\left( \frac{2\pi }{\omega
_{0}}\Gamma \right) \times \frac{1}{2}|\phi _{0}\omega _{0}|^{2}$ .
We define: 
\beq
Q=2\pi \frac{{\cal{E}}_{0}}{{\cal{ E}}_{lost}}=\frac{\omega _{0}}{\Gamma }
\eeq
This holds for an axion RF cavity with $\omega _{0}=m_{a}$.

Much of what limits
$Q$ in an RF cavity is resistive lost, but we can define an accessible signal
power output as: 
\beq
\label{P0}
P_{0}=\epsilon \left( \frac{\omega _{0}}{2\pi }\right) {\cal{E}}_{lost}= \epsilon\frac{ m_{a} {\cal{E}}_0}{Q}
\eeq
Using 
$m_{a}=2\pi \times 1.46
(f_{a}/10^{12})$ 
GHz
$=9.1\times
10^{9}$ Hz,
we have a signal power 
$P_0 = 8.48\times 10^{-22} (2\epsilon)$ watts,
where our numerical coefficient corresponds
to an efficiency of $\epsilon= 1/2$.
\footnote{ A. Chou \cite{Chou2} uses $m_{a}=2\pi  \times  1.0\
\times 10^{9}=6.28\times 10^{9}$  GHz,
 $Q=\left( 1.3\right) \times 10^{5}$ ,  
and a magnetic field derived from the
total magnetic field energy,  $U_{B}=4$ MJ, or
$B_{0}=  7.16$ Tesla, yielding 
$8.4\times 10^{-22}$ watts, consistent
with rescaling our above result.}

How challenging is the extraction of the signal from
an RF cavity?  We think this is challenging.
Note that the signal power can be computed from
the  Poynting vector flux through
 an effective ``aperature,'' \ie, a hole in
the cavity wall of area $A_{out}$ from which a tranversely
polarized signal can be extracted, \eg, a ``bung hole'' in the barrel.
From Maxwell's equations with an an ohmic current in the cavity wall,
one finds  electromagnetic fields
that attentuate over a skin depth $\delta \sim 1/\sqrt{m_a\sigma/2}$,
and which have a small non-zero
Poynting vector that is reduced by a factor of $Q$ 
relative to the cavity energy density.
The radiated power out of the cavity is then:
\beq
\label{P0p}
 P'_0\sim \frac{1}{Q}\frac{{\cal{E}_0}A_{out}}{V}
\eeq
Given these two routes to computing the power,  $P_0$ and $P'_0$,
we can compute the ratio of the aperature area, $A_{out}$
to the surface area, $A_0$, of the cavity in terms
of $\epsilon$.  We 
readily find:
\beq
\frac{A_{out}}{A_0} 
\approx \epsilon 
\eeq
If we choose $\epsilon = 1/2$,
then ${A_{out}}/{A_0} \approx 0.5$ which is large, and implies that extraction
of energy through a physical aperature is limited, since
most of the surface area
of the cavity is in the aperature hole!
We can  extract signal through a
non-perturbative, small aperature
area, of order $A_{out} \sim \pi (\lambda_a/4)^2$, \ie, 
a ``quarter-wave  aperature.''  For the Poynting
vector magnitude eq.(\ref{P0p}) apropos ADMX, 
we then have $P'_0=\pi K (\lambda_a/4)^2 =2.36\times 10^{-24}\eta^2$ 
$(Q/10^6)$ watts, and $\epsilon =4.7\times 10^{-3} $ .  We 
would thus take a significant
hit in the output power with a less perturbative aperature size.

Hence, if appropriate impedance matching of the cavity to
an output receiver can be acheived, with $\epsilon\sim 1/2$, one would be able
to attain power output of the order of $\sim 10^{-21}$ 
from an RF cavity.  The remaining bottleneck is then  probing for
a signal, which requires integrating and analyzing the output 
and searching for a signal/noise excess
for {\em a given physical cavity tuning}.
This can be done on the order of minutes, but then
the cavity must be retuned to a different resonant frequency 
and the signal integration process
repeated.

\vskip 0.25in

{
The signal fields in an RF cavity are driven by the particular solution
to Maxwell's equations with the axion, which induces an oscillating electric field,
$\overrightarrow{E}_{r}= g_a \theta_0 \overrightarrow{B}_0 $, and a vanishing magnetic field
$\overrightarrow{B}_{r}=0$ throughout space, including within the conducting cavity
walls. The particular solution has vanishing Poynting vector and 
cannot propagate, since $\overrightarrow{B}_{r}=0$.
The resonance is determined by the presence of the conducting wall of the cavity.
In this wall the induced electric field $g_a \theta_0 \overrightarrow{B}_0 $ generates a physical current.
This physical current, in turn, stimulates emission of radiation into the cavity, with the  effect that 
oscillating electric and magnetic
modes, $\overrightarrow{E}'_{r}$ and $\overrightarrow{B}'_{r}$, are generated in the cavity. 
The  
net electric field at the wall vanishes, $\overrightarrow{E}_{r}(R)+\overrightarrow{E}'_{r}(R)=0 $.
At resonance the induced fields are $\propto Q$ and the signal energy
density  is $\propto Q^2$.  Finite conductance in the wall
causes a slight phase shift of $\overrightarrow{E}'_{r}$ and $\overrightarrow{B}'_{r}$ away from  $90^o$
by an amount $\propto 1/Q$, leading to a the non-zero time-averaged Poynting vector
of order $Q$. This is what we depend upon for a detectable signal.

Note that the bulk magnetic field, $B_0$, within the internal volume of the cavity, generates
the particular solution there, but this is playing no role in the physics --- only the $\overrightarrow{B}_0$
field in the walls of the cavity is relevant! 
It would be sufficient to have the magnetic field localized only on the wall and not filling the internal volume.
Moreover, this means that if we can apply a large magnetic field in
any conducting material (or dielectric) it will then radiate observable power into the vacuum.
This is the basis of an array radiator we now discuss below.
}

\subsection{Electric Dipole Radiation from Magnets}

We now turn to the radiation produced by 
localized magnetic fields, \ie, induced oscillating dipole moments,
in the cosmic axion field.
Using the  formula of eq.(\ref{power}), 
(where
we include the factor of $\sqrt{4\pi}$ into $\mu_{Bohr}$, 
or, use the CGS  formula, $P= (g_{a}\theta_0\; \mu_{Bohr} m^2_a )^2/3$ directly)  we can estimate
the power emitted by a single electron in free
space, immersed in the cosmic axion field: 
\beq
P_e \sim 5.192 \times 10^{-82}(\eta)^{-4} \;\makebox{watts}.
\eeq
This follows the dipole distribution of eq.(\ref{power0}). 

While this is infinitesimal power, it
will add coherently for an assemblage of many polarized electrons
within the near-zone of the radiation field, \eg, in an approximately a quarter wavelength volume, 
$\sim (\lambda_a/4)^3$, 
which we will define to be a ``unit cell.''  The unit cell
can be viewed as a Dirac $\delta$-function source, and we can construct an array of such sources.

For a mole of polarized electrons comprising a unit cell, within the near-zone, the power becomes
enhanced to:
$P_{Fe} = N_{Avogadro}^2 P_e\sim 1.88\times 10^{-34}(\eta)^{-4} $ watts.
More generally we can assemble many moles of polarized electrons within
the full quarter-wavelength volume to maximize output power. 
The quarter wavelength volume, defined by the axion
wavelength, is $V_0\sim (5.216\; \eta \;cm)^3=1.420\times 10^2 \eta^3$ cm$^3$.
For iron (Fe), of atomic mass $A=56$ gm/mole and  density $\rho \sim 7 $ gm Fe$/$cm$^3$, we can 
therefore assemble $\rho V_0/A=17.74\eta^3$ moles within a quarter wavelength volume.
Assuming one polarized electron per iron atom, the power emitted by such a  
magnet is then of order $(\rho V_0/A)^2P=5.93\times 10^{-32}\eta^2 $ watts.
Note the scaling behavior in $\eta$ is now controlled by overall the 
power per electron, $\eta^{-4}$, times the quarter wave volume squared, $(\eta^{3})^2$ 
for a net scaling $\propto \eta^2$.

Note that the product of the external dipole magnetic field with approximate volume it occupies, 
$B_0 V_0$, can be used
in place of $N \mu_{Bohr}$ in computing the radiation fields.
The Poynting vector is then $K= (12\pi)^{-1}(g_a\theta_0 m_a^2 B_0 V_0)^2$,
valid up to a $V_0$ of order a quarter wavelength.  The results using this
formula are comparable  to the coherent multi-electron
calculation.

Neodymium-ferrite magnets can produce fields up to $\sim 1.4$ Tesla.
A unit cell composed of this material (using $1.4$ Tesla $\times V_0$
in place of  $\mu_{Bohr}N_{Avogadro}N_m$ in the estimate for Fe) 
yields a power of about $P_{Nd}\sim 1.89\times 10^{-30}\eta^{2}$ watts. 
Clearly, there is a significant advantage to larger field strengths. 

Bear in mind that arrays of unit cells spaced external to the near-zone 
 produce a Poynting vector
flux that is enhanced by $N^2$, but this is focused within a solid angle of
$\sim 1/N$, \ie,  the net power output is not coherently enhanced as $N^2$ for
larger arrays, but only as $N$.
We can, nonetheless, 
use an array of $N$ unit cells to enhance the
overall power output of the array $\propto N$ and 
focus the energy into a small solid angle  to  a receiver.

We can also consider a ``unit slab cell.''
This is  plate of conductor lying in the $xy$ plane with a strong magnetic
field that is imbedded and polarized in the plane of the conductor, \eg, in the $\hat{y}$ direction.
Such a plate will radiate coherently in the directions normal to the $xy$ surface,
in the $\hat{z}$ direction,
and the radiation will be electrically polarized in the $\hat{y}$ direction.

In Appendix A we derive the solution for the radiation field
from maxwell's equations.   
The radiation from the $xy$ slab into the $\hat{z}$ zenith direction
is $\overrightarrow{E}_r=-gB_0\theta_0\cos(m_at-kz)\hat{y}$ and 
$\overrightarrow{B}_r=gB_0\theta_0\cos(m_at-kz)\hat{x}$,
and the time averaged Poynting vector
is given by:
\beq
\label{powwow}
P_0 =\frac{1}{8\pi} (gB_0\theta_0)^2A\hat{z}
\eeq
Here $A$ is the surface area of the slab 
and a factor of $1/2$ arises from time averaging the Poynting vector.
Given that the thickness of the conducting surface
exceeds the skin depth, the upward going radiation
matches the $gB_0\theta_0$ particular solution in the skin of
the conductor, which sources the emitted radiation.
Note that here we have assumed a negligible $B_0$ above
the plane of the radiator. If $B_0$ is constant in  $z$ above
the plane of the radiator, then
the radiated power will acquire an interference term
and eq.(\ref{powwow}) will receive an overall factor of $(1-\cos(kz))$.
This oscillatory behavior in space could prove useful in diagnosing a signal.

We will therefore define a ``unit slab cell'' as a $1\;m\times 1\;m$ conducting plate (\eg, copper)
with a $1$ Telsa field in the plane of conductor.  While this may be an unrealistic
construct experimentally, we can use the result to scale to other parameters.
We thus obtain for the unit slab cell, $P_{slab} = 8.27\times 10^{-29}$ $(B_0/1\;T)^2$(Area$/1\;m^2$) watts.

Essentially what we are doing presently is ``opening up'' the RF cavity, and
creating a simple radiator.  We will not be storing energy 
in a cavity, hence $Q=1$ for us. Instead, by constructing
an array of slab cells we would be exposing the
largest possible conducting surface area and coherently
focusing the axion induced emitted radiation toward a  receiver antenna.
which lies at some point in the $\hat{z}$ direction.

Such a plate may not be easy to construct, but we can scale in
area and field strength to match engineering constraints.  
This could be constructed \eg, by placing race-track magnetic
windings wrapped around the plate in the $\hat{x}$ direction;
alternatively, the plate could be segmented into $\hat{x}$ direction strips 
spaced by the solenoids aligned in the $\hat{y}$ direction.
Segmentation of this configuration is tolerable
 provided the average large distance (compared to an axion wavelength)
is the regular square. 

\noindent
In summary:
\bea
\makebox{Single electron} & &  P_e \sim 5.192 \times 10^{-82}(\eta)^{-4} \;\makebox{watts}.
\nonumber \\
\makebox{Fe Unit Cell} & &  P_{Fe} \sim 1.88\times 10^{-34}(\eta)^{-4} \;\makebox{watts}
\nonumber \\
\makebox{$1.4$T Nd Unit Cell} & &  P_{Nd}\sim 1.89\times 10^{-30}\eta^{2} \;\makebox{watts}
\nonumber \\
\makebox{$1m^2$ $1$T slab cell} & &  P_{slab} = 8.27\times 10^{-29} \;\makebox{watts}
\nonumber \\
\eea

{

For magnets, we expect to  encounter complications due
to the conductivity of the material, and the effects of eddy currents
that are generated there.    Most high field magnets are good
conductors at GHz frequencies, and  one might expect
that the electric field will be nullified by the
response in the material
and hence in the near-zone of the radiation field.
However, if one inspects eq.(\ref{Er}) one sees that
the electric field  {\em is vanishing } in the near-zone,
due to a cancellation of the electric dipole radiation with the particular solution
from the oscillating axion field throughout space
(analogous to the cancellation in the wall of the RF cavity). The radiation field
in the near-zone is therefore predominantly due to
the curling magnetic component of eq.(\ref{Br}, \ref{Brnear}).
Therefore, the emitted power results for aggregate magnets
of order a quarter wavelength in size is a collective
effect and localized eddy currents may not be
an appreciable effect.  

One might think that we could simply replace the cylindrical magnets with small solenoids.
Here the problem arises that the solenoidal $B$-field itself can only
generate the non-propagating particular solution in the vacuum, and we require the 
interaction of this field in matter to produce the radiation (this is akin to the RF cavity case).
The solenoidal magnet  is clad with the conducting wire windings.
The magnetic field can only penetrate significantly into this material within
the interior of the solenoid, to generate a small
cavity radiation (far below resonance) within the magnet.
The  external field is weak at
the conducting material surface and does not tend
to lie within the skin depth of the windings there, and we do not
expect significant radiation to be generated by a solenoid itself (this
is equivalent to the fact that a sealed RF cavity tends not to radiate
externally if the return flux at the external conducting surface is small). 
Nonetheless, we could exploit strong field solenoids by arraying them 
 in  the $xy$ plane with a conducting material strategically spanning the inter-magnet space 
(\eg, ``fins'') in the 
$xy$ plane. This will produce a coherent radiation signal in the $z$ direction,
which we discuss below.  The number of unit-cell magnets 
in the array $N$ then becomes equivalent to
the RF cavity $Q$.

Given an xy array of magnets with a polarization in the $\hat{x}$ 
direction, we can compute the discrete sum to obtain the array radiation
fields and Poynting vector.  Alternatively, we can take a 
continuous limit and the array becomes equivalent to
a slab of conductor with an imbedded magnetic field in the  $\hat{x}$ 
direction.  We can imagine using race-track solenoids wound around
the slab, with exposed segments of conductor as the radiation
surfaces.  Other configurations can be investigated.  In the following we estimate the
signal and integration time necessary to detect in this
kind of general xy array configuration. 
}

\subsection{Axion Signal in a 2-D Axion $\rightarrow $ Photon Antenna}

We presently consider a schematic  experimental configuration.  
This is a variation on a scheme proposed by \cite{Ringwald}.
The purpose of this is to provide an example of how one might
attempt to exploit axion induced
electric dipole radiation for detection in a broadband antenna.  This is
a rough ``initial pass,'' 
and we will refine it elsewhere \cite{Hill3}.

Consider a smooth surface (a floor) which we define to be the 
$xy$ plane. In this plane we assume we have placed a number $N$ of the
unit  cells defined in the preceding section. 
The  power emitted into the $\hat{z}$ direction
is then given by:
\beq
\label{slab}
P_{slab} =\frac{1}{8\pi} (g_aB_0\theta_0)^2 A_{total}
\eeq
where $A_{total}=NA$ is the total exposed conducting surface
(and, as discussed above, if $B_0$ is constant
above the plane of the radiator the result becomes 
$ P_{slab} =\frac{1}{8\pi} (g_aB_0\theta_0)^2 A_{total}(1-\cos(kz))$ ).

Details of geometric focusing to a receiver  antenna will be developed elsewhere.
It is not hard to see that the beam of radiation from the $xy$ array
will be coherently focused into a small solid angle and 
can be collected by a parabolic antenna looking
down on the array  (A variation on this \cite{Ringwald} would be to arrange
the slab cells on the surface of a mathematical sphere of radius $R$ and collect the signal at the
focal point of the sphere;  we think it may be advantageous
to use an independent parabolic antenna and a planar array).
The issues of antenna design and optimization are beyond
the scope of our present discussion.  Let us crudely estimate
presently the expected signal sensitivity and baseline requirements.

Quantitatively,  for an extremely optimistic ``benchmark,''
 we'll assume $B_0=10$ Tesla and that we have configured an
array of $10\times 10$ slab cells into an effective $10m\times 10m=10^6\; cm^2$
conducting surface.
We therefore  we have  an emitted power of  $P_{array} \approx 8.27\times 10^{-25} $ watts
that is independent of $\eta$.
This is, as we'll see momentarily, 
more than adequate for detection in a cryogenic environment, and we will
illustrate the scaling with array area, temperature,  and magnetic field
to back down to a more  minimal experimental scale.

The signal received in our antenna will be electronically  filtered 
into a ``pass-band'' that we
take to be of order $\sim 0.5 m_a$ to $\sim 1.5 m_a$
for a hypothetical $m_a= 1.46\times 10^{9} \;\eta^{-1} $ Hz.  The 
pass-band is subdivided into frequency bins $\Delta f$.
These bins are given by the natural linewidth
of the axion itself, defined by the so-called ``axion $Q_a$.'' 
The axion $Q_a \sim 10^6$ arises as a 
line broadening due to motion within the galactic dark matter halo, which
generates fluctuations in the axion velocity,
$\beta\sim 10^{-3}$.  Hence, our frequency bins
are of order  $\Delta f \sim m_a\times \beta^2 \sim 10^3$ Hz,
and we can select $\sim 10^6$ such bins within our
pass-band.
The pass-band signal is recorded by the radio receiver over a long
integration time and Fast Fourier Transformed. The experiment can be repeated
for other pass-bands (other hypothetical $m_a$'s).

The thermal noise power in a bin is given by the (one-dimensional)
Bose-Einstein distribution for thermal photons incident on the 
the receiver.  The noise power in a single frequency bin
is $P_{noise}=2T\Delta f$.  We will assume cooling
of the antenna (and other noise sources) to $T=0.1^oK$ 
corresponding to $P_{noise}=4.02\times 10^{-21}$ watts.
Hence our signal/noise ratio is $P_{array}/P_{noise} = 2.06\times 10^{-4}$.
This is small, but the number of noise photons random walks in the number
of ``samples,'' $t\Delta f  $, hence a signal can be observed with sufficient integration time
to reduce the fluctuations in the noise, $``\Delta N_{noise}" = (P_{noise}/m_a)\sqrt{t\Delta f }$,
below the number of signal events, $``N_{signal}"=(P_0/m_a)t\Delta f$.

This is summarized by the Dicke Radiometer Formula for $t$:
\beq
t = 2\frac{P_{noise}^2}{P_0^2\Delta f} 
\eeq
If we plug in the results for the minimal array we
obtain:
\beq
t= 9\;\makebox{hours}\;
\eeq
We give a tabulation of integration times for various arrays,
each assumed to be 10m $\times$ 10m.  

\vspace{0.2 in}
{\begin{tabular}{||l|c|c|l||}
\hline
Array  & $B_0$  & power   & time  \\ \hline
ND-Fe  & 1.4 Tesla          & $7.6\times 10^{-27}$  watts              & 3.2 years \\ \hline
Solenoids  & 3 Tesla   & $6.8\times 10^{-26}$  watts              & 56 days\\ \hline
10m $\times$ 10m  & 10 Tesla          & $8.27\times 10^{-25}$  watts              & 9 hours\\ \hline
\end{tabular}}

\vspace{0.05in}
{\begin{center}
TABLE II.  Signal integration times for planar arrays of conductors, $10m\times 10m$.
\end{center}}
\vspace{0.2in}

For example, we can use fixed field ND magnets, 
with magnets spaced a half-wavelengths to form a fixed field array.  We  also
consider $3T$ solenoids,  forming a layer under a conducting plane, where
the return flux lies in the plane (various other configurations can be
considered).  Note the scaling in each case is:
\beq
t\propto \left(\frac{B_0 }{B}\right)^4\left(\frac{10m\times 10m}{A}\right)^{2}\left(\frac{T}{0.1^oK}\right)^2
\eeq
For example, a $20m\times 20m$ Nd-Fe magnet array we would have an integration
time reduced by $1/(4)^2$, or $2.4$ months.  Such an array would require $\sim 
4\times 10^4$ magnets of order $\sim 5$ cm in scale length spaced by $\sim 10$ cm.
Note that, despite the choice of quarter wavelength unit cells magnets 
and spacing of order $\lambda_a/2$ for some $\eta\sim 1$, the signal in the
$\hat{z}$ direction is fairly broad-band and  independent
of $\eta\sim \eta_0$.  In Appendix A we describe briefly how the discrete
array power approaches the slab power in the continuum limit.

Note that if the $xy$ array is a slightly spherical surface, then dipole radiation will
be focused to the focal point, located at the mathematical center of the sphere.
The cells would be emitting dipole radiation to the focal point in a common phase
(see \cite{Ringwald}).   The use of  a parabolic reflector over
a flab $xy$ planar array has the  equivalent common phase relationship 
from the a flat  array at the receiver focal point.  Nonetheless,
one might exploit the combination of spherical array
and semi-parabolic receiver antenna.  

With an observed signal
in hand, one would be able to exploit interference effects
such as the $\propto (1-\cos(kz))$ modulation above the array in aconstant $B_0$
field.  It may also be possible
to exploit polarization of the radiation and multiple receivers
to reduce signal to noise and cryogenic
requirements. The signal will be electrically polarized with the
oreintation of the magnets or current, while the noise is unpolarized. 

This is a simple radiating system that does not
depend upon a resonance condition and does not require a 
physical retuning of the array, and is fairly broadband. 
A more detailed discusion of a conceptual 
array-based axion search
experiment will be given elsewhere \cite{Hill3}. 

\section{Conclusion}

Our principal result is that the perturbative
interaction with the cosmic axion, though the anomaly, causes an electron
to acquire an oscillating electric dipole moment.
The result is general: any static magnetic source field in the presence
of the cosmic axion will generate an
oscillating electric field, hence any magnetic moment becomes
an effective electric moment\footnote{As noted in ref.\cite{Hill1}, the
converse is not true,\ie, a static electric field does not become
an oscillating magnetic field, a consequence of the fact that the 
cosmic axion field effectively breaks Lorentz invariance.}.  

As an explicit example, we demonstrate that classically
that a stationary magnetic dipole field radiates as
an oscillating electric dipole field; the near-zone structure is that of a 
time dependent``Hertzian" electric dipole source
of frequency $m_a$.   Likewise, the electron will have an OEDM
that is proportional to its spin, and will experience electric dipole interactions
with applied electric fields.  In addition we have the appearance of the nonlocal term,
the analogue of the transverse current in radiation gauge.
This nonlocal effect is present in {\em any} axion electromagnetic interaction,
such as the case of  fields induced by the axion and applied magnetic
fields in RF cavities (see Section V.).  It does not affect
the oscillating electric dipole interaction of the electron with a source free 
electric  field, \ie as in radiation or cavity modes.

We have obtained an induced 
oscillating  electric dipole moment for the electron,
proportional to  the magnetic moment, $2g_a\theta_0\cos(m_at)\mu_{Bohr}$.
The result is quantitatively $
\approx 3.2\times 10^{-32}\times 10^{-32}(g_a/10^{-3})\cos(m_at)$
e-cm. 
The result is  two orders of magnitude greater
than the typical result expected
for the nucleon,  
$d_N\sim  3.67\times 10^{-35}\cos(m_at)$ e-cm \cite{budkher}, 
about four orders of magnitude
from the  limit on the constant electric dipole moment  of the electron, 
 $d_e\leq 8.7\times 10^{-29}$ e-cm, \cite{ACME}. 

The result is a general low energy theorem and applies to any
static magnetic system.
Axion electromagnetic anomaly effects are essentially local oscillating
 dual rotations that lead to potentially observable signals.
The axion anomaly
perturbs the system by locally producing a physical, infinitesimal,
time dependent dual rotation of 
$\delta\overrightarrow{E} = g_{a\gamma\gamma}\theta_0(t)\overrightarrow{B}$,
effectively rotating a magnetic moment to an oscillating electric dipole moment.
The duality
of axion electrodynamics also implies 
the absence of induced oscillating magnetic moments from static electric dipoles.

The requirement of an explicit appearance of $\partial_t\theta(t)$ 
in physical quantities
is a source of confusion to some people.  This issue is not exclusive to OEDM's,
but also arises  at the classical
level in well-known solutions to axion-Maxwell's equations, such as in conventional  RF cavities.
As shown in Section V.,  in an RF cavity  with an applied classical constant
background field $\vec{B}_0$, an oscillating electric
field develops  that has the form $\vec{E} \propto \tilde\theta(t) \vec{B}_0$.
One does not see the explicit $\partial_t\theta(t)$ in the physical $\vec{E}$,
so one might wonder how it turns off in the $m_a\rightarrow 0 $ limit?
The Maxwell equations for
$\vec{E}$ and $\vec{B}$  are of the linear, first order, inhomogeneous form, 
and must therefore {\em have a boundary condition}.
If one is careful, therefore, one finds that the electric field
is proportional to:
\beq
 \tilde\theta(t)=\int^t_{t_0} d\tau\; \partial_\tau  \theta(\tau)=\theta(t)-\theta(t_0) 
\eeq 
Hence,  $\tilde\theta(t)\rightarrow 0$ when $\partial_t\theta(t)\rightarrow 0$, owing
to the boundary condition.
So, when one measures $\vec{E}(t)$ in
an RF cavity one is only measuring its value relative to an earlier value $\vec{E}(t_0)$. 
In the present case of OEDM's, however, this nonlocality is more subtle,
but in the low energy zero--electron recoil limit of interest it reduces back to $\propto \theta(t)$. 

The existence of such phenomena may
imply a number of potentially sensitive venues.
The collective magnetization of any
subtance, \eg, polarized ferromagnets such as iron or strong rare earth
magnets such as neodymium-ferrites, will acquire  induced
oscillating  electric dipole moments that may
provide accessible signals.

We have presented  estimates for some $xy$ array configurations
that probe the coherent radiation generated by an assemblage
of a large number of polarized electrons or electruic currents. Certainly other
geometries can be considered, including $1D$ and
even $3D$ ``crystal'' arrays.
 These configurations can be
built out of ``unit cells'' which are quarter (axion) wavelength
volumes composed of  magnetic materials.
We find that reasonable laboratory scale configurations, particularly
the 2-D large array, 
may provide detectable coherent power outputs in excess of
$\sim 10^{-25}$ watts. These arrays have signal/noise integration times that 
shrink inversely with increasing magnetic field strength as $\propto B^{-4}$
and area as $\propto A^{-2}$. 
These arrays seem potentially advantageous in probing the axion ``sweet spot''
of $10^{10} \leq f_a\leq 10^{12}$ GeV.
These are very preliminary considerations, but appear worthy of 
further more detailed analysis.

We note that our present paper and \cite{Hill1} has not been without some controversy.
In part, in some earlier versions we mis-stated how the decoupling of the axion works for the OEDM
in the Introduction.
Hoswever, in \cite{Flambaum:2015ica} a calculation was done in a pure Coulomb potential,  which yielded zero
and was used to argue that our overall result is zero. As we have shown explicitly,
the Coulomb potential part of the result is always a total divergence, and indeed vanishes.
The OEDM is intrinsically time dependent and is nonzero as an action. The axion decoupling
has been thoroughly explored in the present paper and
this issue has now been resolved favorably for the present work and \cite{Hill1},
(see \cite{Hill:2015lpa}).  We have explicitly addressed the criticism that
our result did not display decoupling.
in the $m_a\rightarrow 0$ limit.  The decoupling is present, even
in the radiation formula of eq.(\ref{edm00}), but it is not
always manifest, and is subtle in a manner similar to the anomaly
itself, as discussed above in Section III.  

\vspace{0.1in}


\vskip 5 pt
\noindent
{\bf Acknowledgements}

\vspace{0.1in}

I  thank Bill Bardeen for explicitly checking
many of the key theoretical results of  this, and my previous paper,
\cite{Hill1}. 
I especially thank Aaron Chou  and Henry Frisch, also Mike Diesburg, Estia Eichten, Paul LeBrun, Adam Para, William Wester,
and other participants in the Fermilab axion discussion group, for useful discussions.
This work was done at Fermilab, operated by Fermi Research Alliance, 
LLC under Contract No. DE-AC02-07CH11359 with the United States Department of Energy.

\vskip .2in
\noindent
{\bf Appendix A: Slab Radiator Solution }
\vskip .1in
\renewcommand{\theequation}{A.\arabic{equation}}   
\setcounter{equation}{0}  

Here we give the solution for the electric dipole radiation form a conducting
slab with an imbedded magnetic field.
The power from a slab array can be estimated from the
continuum limit of a discrete array of magnets. 
The emitted power from a discrete array of $N$ magnets is
of order:
\beq
 P=g_{a}^{2}\left( \mu
_{Bohr}\right) ^{2}m_{a}^{4}\theta _{0}^{2}N
\eeq 
The axion Compton wavelength is 
$m_{a}\sim \left( 2\pi /\lambda \right) $.
We can identify the magnetic field  in the discrete array
with 
\beq
\overrightarrow{B}_{0}\sim \mu _{Bohr}\left( 2\pi /\lambda \right) ^{3}\sim
\mu _{Bohr}\left( m_{a}\right) ^{3}
\eeq
and the discrete array power is then:
\bea
P & \sim & 
g_{A }^{2}\left( \overrightarrow{B}_{0}\right)
^{2}m_{a}^{-2}\theta _{0}^{2}N 
\nonumber \\
& \sim &  g_{A }^{2}\left( 
\overrightarrow{B}_{0}\right) ^{2}\theta _{0}^{2}\left( \lambda /2\pi
\right) ^{2}N
 \sim  g_{A }^{2}\left( \overrightarrow{B}_{0}\right) ^{2}\theta
_{0}^{2}\times \makebox{(area)}
\nonumber \\
\eea
Let us derive this result carefully  from Maxwell's equations.
The sourced Maxwell equation is:
\beq
\overrightarrow{\nabla }\times \overrightarrow{B}_{r}-\partial _{t}%
\overrightarrow{E}_{r}=g\overrightarrow{B}_{0}\partial _{t}\theta +%
\overrightarrow{J}
\eeq
where $\overrightarrow{B}_{0}$ is a constant background magnetic field; $\theta $ is oscillating axion field and 
$\overrightarrow{E}_{r}$ and $\overrightarrow{B}_{r}$ are oscillating 
response fields.

We assume we have vacuum for  $x<0$  and a conducting medium, such as copper,
 for $x>0$.
$\overrightarrow{B}_{0}$ and  $\theta $ permeate the vacuum and the conductor.
The magnetic field, $\overrightarrow{B}_{0}=B_{0}\widehat{z}$ is parallel to the 
plane of the conductor.
We assume Ohm's Law in the conductor 
\beq
\overrightarrow{J}=-\sigma 
\overrightarrow{E}
\eeq
Introduce a vector potential in radiation gauge for the response fields:
\beq
\overrightarrow{E}_{r}=-\partial _{t}\overrightarrow{A}\qquad \ B_{r}=%
\overrightarrow{\nabla }\times \overrightarrow{A}\qquad \overrightarrow{%
\nabla }\cdot \overrightarrow{A}=0
\eeq
hence:
\beq
\partial _{t}^{2}\overrightarrow{A}-\nabla ^{2}\overrightarrow{A}=g%
\overrightarrow{B}_{0}\partial _{t}\theta +\overrightarrow{J}
\qquad x>0
\eeq
\beq
\partial _{t}^{2}\overrightarrow{A}-\nabla ^{2}\overrightarrow{A}=g%
\overrightarrow{B}_{0}\partial _{t}\theta \qquad x<0
\eeq
The axion induced  particular solution is: 
\beq
\overrightarrow{E}_{0}=-g\overrightarrow{B}_{0}\theta =-\partial _{t}%
\overrightarrow{A}_{0}\;\; \makebox{hence,}\;\;  \partial _{t}^{2}%
\overrightarrow{A}_{0}=g\overrightarrow{B}_{0}\partial _{t}\theta 
\eeq
Note $\overrightarrow{\nabla }\times \overrightarrow{E}_{r0}=g
\overrightarrow{\nabla }\times \overrightarrow{B}_{0} =0$  since 
$ \overrightarrow{B}_{0}$ is assumed constant
(or source free).  Hence,  $\nabla ^{2}\overrightarrow{A}_{0}=0$ and $\overrightarrow{A}%
_{0}$ is non-propagating and constant in space (curl-free).

It is convenient to complexify
the fields and identify the physical fields with the real parts.
We write $\ \theta =>\theta _{0}\exp \left( im_{a}t\right) ,$ then$\ g%
\overrightarrow{B}_{0}\partial _{t}\theta =im_{a}g\overrightarrow{B}%
_{0}\theta _{0}\exp \left( im_{a}t\right) $, 
and $\ \overrightarrow{A}_{0}=-ig\overrightarrow{B}_{0}\theta
_{0}/m_{a}\exp \left( im_{a}t\right) $.
The equation of motion becomes:
\bea
\partial _{t}^{2}\overrightarrow{A}-\nabla ^{2}\overrightarrow{A}%
& = & im_{a}g\overrightarrow{B}_{0}\theta _{0}\exp \left( im_{a}t\right) +\sigma
\partial _{t}\overrightarrow{A}\;\;  (x>0)
\nonumber \\
\partial _{t}^{2}\overrightarrow{A}-\nabla ^{2}\overrightarrow{A}
 & = & im_{a}g%
\overrightarrow{B}_{0}\theta _{0}\exp \left( im_{a}t\right) \;\; (x<0)
\eea
$\overrightarrow{A}$  is composed of homogenous propagating and
non-propagating solutions. Define:
\bea
\overrightarrow{A} & = & \overrightarrow{P}e^{im_{a}t} +\overrightarrow{H}e^{im_{a}t+k^{\prime}x} 
-\frac{ig\overrightarrow{B}_{0}\theta _{0}}{m_{a}}e^{im_{a}t} \;\; (x>0)
\nonumber \\
\overrightarrow{A}& = & \overrightarrow{K}e^{im_{a}t+kx}\ -\frac{ig\overrightarrow{B}%
_{0}\theta _{0}}{m_{a}}e^{im_{a}t}  \;\;\;\;\; (x<0)
\eea
We  introduce a  parameter:  $\gamma =m_{a}/\sigma $
and we will work in the limit of large conductance,  $\gamma <<1.$
Substituting into the equations of motion we have:
\bea
-\left( m_{a}^{2}+im_{a}^{2}/\gamma \right) \overrightarrow{P}& = & g\overrightarrow{B}_{0}\theta
_{0}({m_{a}}/{\gamma })
\nonumber \\
-\left( m_{a}^{2}+im_{a}^{2}/\gamma -k^{\prime 2}\right) \overrightarrow{H}& = & 0
\nonumber \\
-\left( m_{a}^{2}-k^{2}\right) \overrightarrow{K}&=& 0\qquad 
\eea
Hence: 
\bea
\overrightarrow{P} & =& \frac{ig\overrightarrow{B}_{0}\theta _{0}}{m_{a}\left( 1-i\gamma \right) }
\nonumber \\ 
 k^{\prime }& = &\pm m_{a}\sqrt{1+i/\gamma }
\nonumber \\
k & =& m_{a}
\eea
The sign of $k^{\prime }$ is determined by the asymptotic boundary condition
that
$\overrightarrow{H}\rightarrow 0$ as $x\rightarrow \infty .$
The boundary conditions at $x=0$ match the fields and 
their spatial derivatives on L and R: 
\bea
\frac{ig\overrightarrow{B}_{0}\theta _{0}}{m_{a}\left( 1-i\gamma
\right) }+\overrightarrow{H} & = & \overrightarrow{K}
\nonumber \\
\sqrt{1+i/\gamma }\overrightarrow{H} & = & \overrightarrow{K}
\eea
whence:
\bea
\overrightarrow{H} & = & \frac{ig\overrightarrow{B}_{0}\theta _{0}}{m_{a}\left(
1-i\gamma \right) \left( \sqrt{1+i/\gamma }-1\right) }
\nonumber \\
& &  \approx \sqrt{\gamma }%
\frac{ig\overrightarrow{B}_{0}\theta _{0}}{m_{a}}e^{-i\pi /4}+\gamma \frac{g%
\overrightarrow{B}_{0}\theta _{0}}{m_{a}}
\nonumber \\
\overrightarrow{K} & = &  \sqrt{1+i/\gamma }H\approx \left( \frac{ig\overrightarrow{B}%
_{0}\theta _{0}}{m_{a}}+\sqrt{\gamma }\frac{ig\overrightarrow{B}_{0}\theta
_{0}}{m_{a}}e^{i\pi /4}\right) 
\nonumber \\
\overrightarrow{P} & = & \frac{ig\overrightarrow{B}_{0}\theta _{0}}{m_{a}\left( 1-i\gamma \right) }
\approx \frac{ig\overrightarrow{B}_{0}\theta _{0}}{m_{a}}-\frac{\gamma g%
\overrightarrow{B}_{0}\theta _{0}}{m_{a}}
\eea
We therefore have the resulting solution for small $\gamma <<1$:
\bea
\overrightarrow{A}& \approx & 
\left( \sqrt{\gamma }e^{i\pi /4}+\gamma \right) 
\frac{g\overrightarrow{B}_{0}\theta _{0}}{m_{a}}
e^{im_{a}t+i\frac{m_{a}}{\sqrt{2\gamma }}x-
\frac{m_{a}}{\sqrt{2\gamma }}x }
\nonumber \\ 
& & \qquad \qquad \qquad \qquad
-\gamma\frac{ g\overrightarrow{B}_{0}\theta_{0}}{m_{a}} e^{im_{a}t}\qquad  (x>0)
\nonumber \\
\overrightarrow{A} & \approx &  \frac{ig\overrightarrow{B}_{0}\theta _{0}}{m_{a}}%
\left( 1+\sqrt{\gamma }e^{i\pi /4}\right) e^{(im_{a}\left( t+x\right) )}
\nonumber \\ 
& & \qquad \qquad  -
\frac{ig\overrightarrow{B}_{0}\theta _{0}}{m_{a}}e^{\left( im_{a}t\right)
}\qquad (x<0)
\eea
We thus see, in the $\gamma\rightarrow 0$  limit, the overall amplitude of  $\overrightarrow{A}\rightarrow 0$ 
for $x>0$, and the propagating component attenuates into the conductor with a skin depth  of
$1/\sqrt{m_{a}\sigma }.$  For  $x<0$, in the $\gamma\rightarrow 0$  limit, we see that the
induced propagating plane wave
has an overall amplitude that has become locked to 
the magnitude of the  non-propagating axion induced particular solution.
For  $x=0$ w see that the solution approaches the
usual conducting boundary condition, $ \overrightarrow{E}(x=0)=0$

The main point is to ilustrate that
 a conductor converts the non-propagating particular solution 
into detectable radiation.  This is how the walls of the RF cavity pump radiation into the
volume; the magnetic field in the empty volume plays no
role, only its permeating the walls induces the signal.
This also illustrates how the slab unit cell acts as a radiation
source.

\vskip 0.5 in
 


\begin{thebibliography}{99}
\bibitem{Hill1} 
  C.~T.~Hill,
  Phys.\ Rev.\ D {\bf 91}, 111702 (2015); note: in
this paper we used $g_a=10^{-3}$, while in the present paper
we have adopted the normalization  $g_a=2.3\times 10^{-3}$,
as in ref.\cite{KSVZ}.
\end{thebibliography}
\end{document}